\documentclass[a4paper,11pt]{article}
\usepackage[utf8]{inputenc}

\usepackage[margin=2.5cm]{geometry}

\usepackage{amsmath}
\usepackage{amsfonts}
\usepackage{graphicx}
\usepackage{caption}
\usepackage{subcaption}
\usepackage{float}
\usepackage{comment}

\usepackage{authblk}
\usepackage{amssymb}

\usepackage{epstopdf}
\usepackage{epsfig}
\usepackage{hyperref}

\usepackage{color}
\usepackage[normalem]{ulem}
\usepackage{xcolor}

\allowdisplaybreaks[1]

\def \nn {\nonumber} 

\begin{document}

\title{Universal relations for quasinormal modes of \\ neutron stars in $R^2$ gravity.}
\author[1]{Jose Luis Bl\'azquez-Salcedo \thanks{\href{mailto:jose.blazquez.salcedo@uni-oldenburg.de}{jlblaz01@ucm.es}}}
\author[1]{Luis Manuel Gonz\'alez-Romero \thanks{\href{mailto:mgromero@ucm.es}{mgromero@ucm.es}}}
\author[2]{Fech Scen Khoo \thanks{\href{mailto:fech.scen.khoo@uni-oldenburg.de}{fech.scen.khoo@uni-oldenburg.de}}} 
\author[2]{Jutta Kunz \thanks{\href{mailto:jutta.kunz@uni-oldenburg.de}{jutta.kunz@uni-oldenburg.de}}} 
\author[2]{Vincent Preut \thanks{\href{mailto:vincent.preut@uni-oldenburg.de}{vincent.preut@uni-oldenburg.de}}} 
\affil[1]{Departamento de F\'isica Te\'orica and IPARCOS, Facultad de Ciencias F\'isicas, 
Universidad Complutense de Madrid, Spain}
\affil[2]{Institute of  Physics, University of Oldenburg, Postfach 2503, D-26111 Oldenburg, Germany}

\date{\today}

\maketitle

\begin{abstract}
We construct quasinormal modes for neutron stars in $R^2$ gravity in the Einstein frame, considering scalar masses in the sub-neV range. %
In particular, we investigate the fundamental quadrupole fluid f-modes and the dipole fluid F-modes.
Employing six equations of state covering matter content with nucleons, hyperons and quarks, we then propose universal relations for the quadrupole f-modes and dipole F-modes.
The dipole F-modes are ultra-long lived and, for the lower scalar masses, their frequencies are inversely proportional to the corresponding Compton wavelength.
\end{abstract}

\section{Introduction}

Forming stable highly compact objects,
matter in neutron stars is balancing the tremendous gravitational force with the help of repulsive nuclear forces against collapse.
The extreme conditions existing in neutron stars make them outstanding laboratories, to learn about gravity on the one hand and particle physics on the other hand.
On the particle physics side the equation of state (EOS) of nuclear matter in neutron stars is still unknown, although recent years have seen much progress (see e.g.,\cite{Lattimer:2021emm}).
At the same time neutron star properties are being explored in numerous alternative theories of gravity that are motivated by cosmology and quantum gravity \cite{Faraoni:2010pgm,Berti:2015itd,CANTATA:2021ktz}.
\\
The observations of gravitational waves in merger events involving neutron stars and the associated multi-messenger observations are providing an unprecedented wealth of data \cite{LIGOScientific:2016aoc,LIGOScientific:2017vwq,LIGOScientific:2017ync,LIGOScientific:2020aai}.
Such merger events are consisting of three phases: inspiral, merger and ringdown, where the gravitational waves emitted in the ringdown phase of the newly formed highly excited compact object comprise its quasinormal modes (QNMs) (see e.g.~\cite{Andersson:1996pn,Andersson:1997rn,Kokkotas:1999bd,Berti:2015itd}).
These represent characteristic modes of gravitational radiation from the object, encoding its properties and its reaction to perturbations.
QNMs are also sensitive to the nuclear EOS and to the gravitational force, of course.
Thus the study of QNMs can, in principle, strongly enhance the current knowledge of neutron stars and gravity, once sufficiently accurate data is provided by future gravitational wave detectors.
\\
While at first a caveat seems to be a possible degeneracy of effects arising from the EOS dependence of the QNMs, the use of \textit{universal relations} allows to reduce the EOS dependence and thus lift the degeneracy to a large extent (see the reviews  \cite{Yagi:2016bkt,Doneva:2017jop}).   
Universal relations are typically obtained for adequately scaled dimensionless physical quantities, such that when considered for a large set of EOSs relations between properties hold, that retain very little EOS dependence.
Since universal relations arise both in general relativity and in alternative theories of gravity, deviations of these relations may betray the proper gravity theory, or at least yield constraints on the coupling parameters of these theories.
\\
Universal relations of QNMs were first noticed and studied in general relativity \cite{Andersson:1996pn,Andersson:1997rn,Benhar:1998au,Benhar:2004xg,Tsui:2004qd,Lau:2009bu,Blazquez-Salcedo:2012hdg,Blazquez-Salcedo:2013jka,Chirenti:2015dda,Lioutas:2021jbl,Sotani:2021nlx,Sotani:2021kiw,Zhao:2022tcw}.
They have been shown to exist for axial and polar quadrupole ($l=2$) modes including the fundamental (f) mode, the first pressure (p) mode, and the first curvature (w) mode of static neutron stars.
For axial modes, universal relations have also been studied in some alternative theories of gravity \cite{Blazquez-Salcedo:2015ets,Blazquez-Salcedo:2018tyn,Blazquez-Salcedo:2018qyy,AltahaMotahar:2018djk,Blazquez-Salcedo:2018pxo,AltahaMotahar:2019ekm}.
Axial modes are pure gravity modes, and there is no coupling to matter or possible scalar degrees of freedom, reducing the complexity of the axial perturbation equations.
Polar modes, on the other hand, couple to matter as well as scalar fields, when present in alternative theories of gravity.
This leads to highly involved sets of perturbation equations, unless the Cowling approximation is used, where the perturbations of the spacetime and the gravitational scalar field are frozen (see e.g., \cite{Sotani:2004rq,Staykov:2015cfa}).
Clearly, the full set of perturbations equations results in a much richer spectrum of QNMs, including scalar monopole and dipole radiation as well as scalar-led quadrupole and higher $l$ modes \cite{Sotani:2014tua,Mendes:2018qwo,Blazquez-Salcedo:2020ibb,Kruger:2021yay,Dima:2021pwx,Blazquez-Salcedo:2021exm,Blazquez-Salcedo:2022dxh}.
\\
Among the alternative theories of gravity, $f(R)$ theories have received much attention \cite{Sotiriou:2008rp,DeFelice:2010aj,Capozziello:2011et}.
Here a rather attractive theory is $R^2$ gravity with $f(R) = R + a R^2$, and coupling constant $a$. 
Whereas $f(R)$ theories are formulated in the physical Jordan frame, they are more amenable to detailed neutron star analysis after transformation to the Einstein frame corresponding to a scalar-tensor theory, as formulated for $R^2$ gravity by Yazadjiev et al.~\cite{Yazadjiev:2014cza,Staykov:2014mwa} (see, however, \cite{Astashenok:2017dpo}).
The coupling constant $a$ is then simply related to the mass $m_\phi$ of the gravitational scalar field. 
Due to their relative simplicity, axial modes and some of the their associated universal relations have been analyzed first for $R^2$ gravity \cite{Blazquez-Salcedo:2018qyy}.
The analysis of the full set of polar modes has only recently been initiated, with calculations of monopole ($l=0$) and quadrupole ($l=2$) modes \cite{Blazquez-Salcedo:2020ibb,Blazquez-Salcedo:2021exm}.\\
The monopole modes represent radial modes.
In general relativity the radial matter ($F$, $H_i$) modes correspond to normal modes, since monopole radiation is not allowed.
Moreover, analysis of the radial modes beyond the maximum mass of the neutron stars reveals their instability.
When a scalar degree of freedom is present, however, monopole radiation arises, manifesting itself in $R^2$ gravity in the form of ultra-long lived radial matter (F) modes as well as in additional scalar ($\phi$) modes \cite{Blazquez-Salcedo:2020ibb,Blazquez-Salcedo:2021exm}.
The scale of the frequency of the radial matter modes is set by the size of the star for the $F$-mode (and its multiples for the $H$-modes), when the Compton wavelength of the scalar field is small as compared to size of the star.
Otherwise, the scale is set by the scalar field mass, with the $F$-mode following roughly the frequency dependence of the $\phi$-mode \cite{Blazquez-Salcedo:2020ibb,Blazquez-Salcedo:2021exm}.
\\
Here we extend our previous analysis and obtain also the lowest polar dipole ($l=1$) F-modes. Moreover, we present to our knowledge for the first time universal relations for polar dipole and quadrupole modes in an alternative theory of gravity, beyond the Cowling approximation.
We have organized the paper as follows.
In section 2 we present the theoretical settings, recalling the action of nuclear matter coupled to $R^2$ gravity in both the Einstein and the Jordan frame.
We then discuss the background equations for the neutron stars, including slow rotation, and present the polar perturbation equations.
Next we present our results in section 3, starting with a discussion of the neutron star mass, radius and moment of inertia.
Then the spectrum of quadrupole f-modes and their universal relations are discussed.
Finally, the spectrum and universal relations for the dipole F-mode are presented.
We give our conclusions in section 4. The Appendix shows the set of the explicit $l=1$ equations (\ref{Appendix_eqs_l1}), and exhibits the Tables with the fit parameters for the universal relations (\ref{Appendix_tables_l2} and \ref{App_dipole}).

\section{Theoretical Setting}

\subsection{Neutron stars in $R^2$ gravity}

We consider the scalar-tensor theory action 
in the Einstein frame ($G=c=1$) \cite{Yazadjiev:2014cza,Staykov:2014mwa}, 
\begin{equation}
S [g_{\mu\nu},\phi] = \frac{1}{16\pi } \int d^4x \sqrt{-g}
\big( R - 2\partial_{\mu}\phi \, \partial^{\mu}\phi 
- V(\phi) + L_{M}(A^2(\phi)g_{\mu\nu},\chi) \big)~,
\label{EinsteinAction}
\end{equation}
with the Brans-Dicke coupling function
\begin{equation}
    A(\phi)= e^{-\frac{1}{\sqrt{3}}\phi} 
\end{equation}
in the matter action $L_M$,
and the potential term
\begin{equation}
V=\frac{3m_{\phi}^2}{2} \big(1- e^{-\frac{2\phi}{\sqrt{3}}}\big)^2 \ .
\label{pots}
\end{equation}
The Einstein frame action (\ref{EinsteinAction}) is equivalent to the {$R^2$ gravity} action in the Jordan frame 
\begin{equation}
S[g_{\mu\nu}^*] = 
\frac{1}{16\pi} \int d^4x  \sqrt{-g^*} \big( R^* +  {a} {R^*}^2 + L_{M} (g_{\mu\nu}^*, \chi) \big)
~.
\label{fR}
\end{equation} 
The theory parameter ${a}$ is related to the scalar field mass,
\begin{equation}
    m_{\phi} = \frac{1}{\sqrt{6 {a}}} \ .
\end{equation}

From the action (\ref{EinsteinAction}), the field equation for the metric $g_{\mu\nu}$ is given by,
\begin{equation}
G_{\mu\nu} = T^{(S)}_{\mu\nu} + 8 \pi T^{(M)}_{\mu\nu}
-\frac{1}{2}V(\phi)g_{\mu\nu}
~,
\label{eq_G}
\end{equation}
with Einstein tensor $G_{\mu\nu} = R_{\mu\nu} - \frac{1}{2}Rg_{\mu\nu}$.
The energy-momentum tensor for the scalar field is
\begin{equation}
T^{(S)}_{\mu\nu}=2\partial_{\mu}\phi\partial_{\nu}\phi -
g_{\mu\nu} \partial^{\sigma}\phi\partial_{\sigma}\phi
\ ,
\end{equation}
and for the matter, it is
\begin{equation}
T^{(M)}_{\mu\nu} = (\rho + p)u_{\mu}u_{\nu} + pg_{\mu\nu}
~,
\label{T_matter}
\end{equation}
where the pressure {$p$} and density {$\rho$} in the Einstein frame are related to the {respective physical} Jordan frame quantities $\hat{p}$ and $\hat{\rho}$ via the coupling function $A$,
\begin{equation}
p = A^4 \hat{p}~, \quad \rho = A^4 \hat{\rho} ~.
\label{p_rho_EJ}
\end{equation}
The dynamical field equation {for the scalar field following} from action (\ref{EinsteinAction}) is
\begin{equation}
\nabla_{\mu}\nabla^{\mu}\phi = -4\pi\frac{1}{A}\frac{dA}{d\phi} T^{(M)} + \frac{1}{4} \frac{dV}{d\phi}
~.
\label{scalar_eom}
\end{equation}

\subsection{Background ansatz and properties}

The metric is {chosen} static and spherically symmetric
\begin{equation}
ds^2 = g_{\mu\nu}^{(0)} dx^{\mu} dx^{\nu} = -e^{2\nu(r)} dt^2
+ e^{2\lambda(r)} dr^2 + r^2 (d\theta^2 + 
\text{sin}^2 \theta \, d\varphi^2 )
~,
\end{equation}
the scalar field, energy density and pressure are then functions of $r$,
\begin{eqnarray}
\phi=\phi_0(r) ~, \hat{\rho}=\hat{\rho}_0(r) ~, \hat{p}=\hat{p}_0(r) ~,
\end{eqnarray}
and the four-velocity of the static fluid 
is given by
\begin{eqnarray}
u^{(0)}=-e^{\nu}dt ~.
\end{eqnarray}

The equations for the static functions inside the star are {then given by}
\begin{eqnarray}
&&\frac{1}{r^2}\frac{d}{dr}\left[r(1- e^{-2\lambda})\right]= 8\pi
A_0^4 {\hat{\rho}_0} + e^{-2\lambda}\left(\frac{d\phi_0}{dr}\right)^2
+ \frac{1}{2} V_0 ~,  \label{eq_lambda} \\
&&\frac{2}{r}e^{-2\lambda} \frac{d\nu}{dr} - \frac{1}{r^2}(1-
e^{-2\lambda})= 8\pi A_0^4 \hat{p}_0 +
e^{-2\lambda}\left(\frac{d\phi_0}{dr}\right)^2 - \frac{1}{2}
V_0 ~, \label{eq_nu}
\\
&&\frac{d\hat{p}_0}{dr}= - (\hat{\rho}_0 + \hat{p}_0) \left(\frac{d\nu}{dr} +\frac{1}{A_0}\frac{dA_0}{d\phi_0}\frac{d\phi_0}{dr} \right) ~, \label{eq_pJ} 
\\
&&\frac{d^2\phi_0}{dr^2} + \left(\frac{d\nu}{dr} -
\frac{d\lambda}{dr} + \frac{2}{r} \right)\frac{d\phi_0}{dr}= 4\pi\frac{1}{A_0}\frac{dA_0}{d\phi_0} A_0^4(\hat{\rho}_0-3\hat{p}_0)e^{2\lambda} + \frac{1}{4}
\frac{dV_0}{d\phi_0} e^{2\lambda} ~, \label{eq_phi} 
\end{eqnarray}
where $A_0=A(\phi_0)$ and $V_0=V(\phi_0)$.

To calculate the moment of inertia at first order in the slow rotation approximation, we introduce the angular velocity of the star $\Omega$ and the inertial dragging $\omega(r)=\Omega-w(r)$ \cite{Hartle:1967he,Sotani:2012eb},
\begin{equation}
ds^2 = -e^{2\nu(r)} dt^2
+ e^{2\lambda(r)} dr^2 + r^2 (d\theta^2 + 
\text{sin}^2 \theta \, d\varphi^2 )-2(\Omega-w)r^2 \text{sin}^2 \theta \, dt \, d\varphi
~.
\end{equation}
The energy density and pressure do not change, but the four-velocity is
\begin{eqnarray}
u=-e^{\nu}(dt+r^2\sin^2{\theta} \, w \, d\varphi) ~,
\end{eqnarray}
and the resulting equation for $w$ is
\begin{equation}
    \frac{e^{\nu-\lambda}}{r^4}\frac{d}{dr}\left[ e^{-(\nu+\lambda)}r^4\frac{dw}{dr}\right]= 16\pi
A_0^4 {(\hat{\rho}_0+\hat{p}_0)}w \ .
\end{equation}
The asymptotic behaviour {of the metric and scalar functions is given by}
\begin{eqnarray}
e^{2\nu}=e^{-2\lambda}\sim 1-2M/r ~, \\
w \sim \frac{2J}{r^3} ~, \\
\phi_0 \sim \frac{1}{r}e^{-m_\phi r} ~,
\end{eqnarray}
where $M$ is the star mass, and $J$ its angular momentum. {Thus} $I=J/\Omega$ is its moment of inertia.

Regularity at the center {of the star requires}
\begin{eqnarray}
\nu(0) = \nu_c ~, \ \ \lambda(0) = 0 ~, \ \ \phi_0(0) = \phi_c ~, \\
\hat{p}(0)=\hat{p}_c ~, \ \ \hat{\rho}(0)=\hat{\rho}_c ~, \\
w(0) =0 ~.
\end{eqnarray}
The border of the star is {located at} $r=R$, where $\hat{p}(R)=\hat{\rho}(R)=0$.

\subsection{Polar perturbations}

Here we focus on polar perturbations {(see e.g. \cite{Regge:1957td,Zerilli:1970se,Thorne:1967a,Price:1969,Thorne:1969rba,Campolattaro:1970,Thorne:1980ru,Detweiler:1985zz,Chandrasekhar:1991fi,Chandrasekhar:1991,Chandrasekhar:1991_,Ipser:1991ind,Kojima:1992ie} for derivation of the appropriate decomposition and ansatz)} and, in particular, on dipolar and quadrupolar perturbations. 
The zeroth order is given by the static and spherically symmetric background.
Introducing the perturbation parameter $\epsilon<<1$, we perturb the background metric, the scalar field and the fluid as follows
\begin{eqnarray}
g_{\mu\nu} = g_{\mu\nu}^{(0)}(r) + \epsilon h_{\mu\nu}(t,r,\theta,\varphi)~, \\
\label{Phiperturb}
\phi = \phi_{0}(r) + \epsilon \delta\phi (t,r,\theta,\varphi)~, \\
\rho = \rho_0 (r) + \epsilon \delta\rho(t,r,\theta,\varphi)~,  \\
p = p_0(r) + \epsilon \delta p(t,r,\theta,\varphi)~, \\
u_{\mu} = u^{(0)}_{\mu} (r) + \epsilon \delta u_{\mu} (t,r,\theta,\varphi)
~.
\end{eqnarray}

The polar metric perturbations can be written as
\begin{equation}
h_{\mu\nu}^{(\text{polar})} = \sum\limits_{l,m}\,\int    
\left[
\begin{array}{c c c c}
r^l e^{2\nu} H_0 Y_{lm} & -i \omega r^{l+1} H_1 Y_{lm} & 
0 & 0 \\
-i \omega r^{l+1} H_1 Y_{lm} &  r^l e^{2\lambda} H_2 Y_{lm} & 
0 & 0 \\
0 & 0
& r^{l+2} K  Y_{lm} & 0
\\
0 & 0  & 
0 & r^{l+2}  \sin^2\theta K  Y_{lm} \\
\end{array}
\right]
e^{-i\omega t} d\omega
~,
\end{equation}
{where $Y_{lm}$ are the spherical harmonic functions.}
The scalar field and fluid perturbations read
\begin{equation} 
\delta \phi =  \sum\limits_{l,m}\,\int  r^l \phi_1 \, 
Y_{lm} e^{-i\omega t} d\omega ~,
\end{equation}

\begin{equation}
\delta \rho = \sum\limits_{l,m}\,
\int   r^l E_{1 } Y_{lm} e^{-i\omega t }d\omega~, \quad
\delta p = \sum\limits_{l,m}\,
\int   r^l \Pi_{1}  Y_{lm} e^{-i\omega t} d\omega~, 
\end{equation}

\begin{eqnarray}
\delta u_{\mu} = 
\sum\limits_{l,m}\,\int    
\left[
\begin{array}{c}
\frac{1}{2} r^l e^{\nu} H_0 Y_{lm}  \\
r^l i\omega e^{-\nu} 
\left(e^{\lambda}W/r -r H_1 \right) Y_{lm}  \\
-i\omega r^l e^{-\nu} V \partial_{\theta} Y_{lm}
\\
-i\omega r^l e^{-\nu} V \partial_{\varphi} Y_{lm}  \\
\end{array}
\right]
e^{-i\omega t} d\omega
~.
\end{eqnarray}
We have introduced the complex frequency $\omega=\omega_R+i\omega_I$ of the QNMs, {where $\omega_R$ represents the characteristic frequency and $\omega_I$ represents the decay rate of the mode.} 
Outside the star there is no fluid, meaning $\Pi_1=E_1=0$, so the perturbations simplify considerably.

Substituting this ansatz into the field equations, it is possible to reduce the problem to a set of ordinary differential equations. 
These equations can be found, for the general case, in \cite{Blazquez-Salcedo:2020ibb,Blazquez-Salcedo:2021exm}.
In this paper we will focus on $l=1$ (dipole) and $l=2$ (quadrupole) perturbations.
{The resulting set of equations for $l=1$ is shown in Appendix \ref{Appendix_eqs_l1}.} 

The procedure to obtain the QNMs is similar to the one followed in \cite{Blazquez-Salcedo:2020ibb,Blazquez-Salcedo:2021exm}: we must impose outgoing-wave conditions as $r\to\infty$, while at the center and at the surface of the star, the perturbation functions must be regular. These translate into a number of conditions that can be imposed numerically (see also \cite{Blazquez-Salcedo:2018pxo} for a description of the general method employed to calculate the modes).


\section{Results}

In this section we present our results. As discussed above, the field equations have to be supplemented by an equation of state {$\hat{\rho}(\hat{p})$ (usually given in the Jordan frame)}. 
Here we choose the following set of realistic EOSs:
\begin{itemize}
\itemsep=-1pt
\item Two EOSs containing plain nuclear matter: SLy \cite{Douchin:2001sv} and APR4 \cite{Akmal:1998cf}. 
\item Two EOSs containing nucleons and hyperons: GNH3 \cite{Glendenning:1984jr}, H4 \cite{Lackey:2005tk}.
\item Two EOSs containing hybrid nuclear-quark matter: ALF2 \cite{Alford:2004pf}, WSPHS3 \cite{Weissenborn:2011qu}.
\end{itemize}
{We also consider}
a simple non-relativistic polytropic equation of state $p=\kappa\rho^{5/3}$, with $\kappa = 0.53802 \cdot 10^{10}$ dyn/(g/cm$^3$)$^{5/3}$ used {for dipole F-modes} \cite{Lindblom:1989} {(see Appendix \ref{App_dipole})}.

{We perform a detailed QNM analysis for a set of $R^2$ theories with coupling constants $a$ satisfying the observational bound from Gravity Probe B, $a\lesssim 5 \cdot 10^{11}$ m$^2$ \cite{Naf:2010zy}.
In terms of the scalar mass this bound translates into $m_\phi \gtrsim 10^{-4}$ neV (see also \cite{Yazadjiev:2014cza}).
In particular, we choose for the scalar field mass the values $m_\phi=0.0108$ neV, 0.0343 neV, 0.0766 neV, 0.1084 neV and 0.3428 neV.
These values belong to the physically interesting mass range \cite{Naf:2010zy,Ramazanoglu:2016kul,Brito:2017zvb}, and at the same time cover well the full range of QNM values for the fundamental quadrupole mode, ranging basically between the limiting theories, general relativity on the one hand and a massless Brans-Dicke theory on the other hand.
As already earlier noted, for $m_\phi=1.08$ neV the fundamental modes approach those of general relativity very closely, whereas $m_\phi = 1$ peV the modes are very close to those of the massless case \cite{Blazquez-Salcedo:2018qyy,Blazquez-Salcedo:2020ibb,Blazquez-Salcedo:2021exm}.
}

\subsection{Mass, radius and moment of inertia}

\begin{figure}[t!]
	\centering
	\includegraphics[width=.32\textwidth, angle =-90]{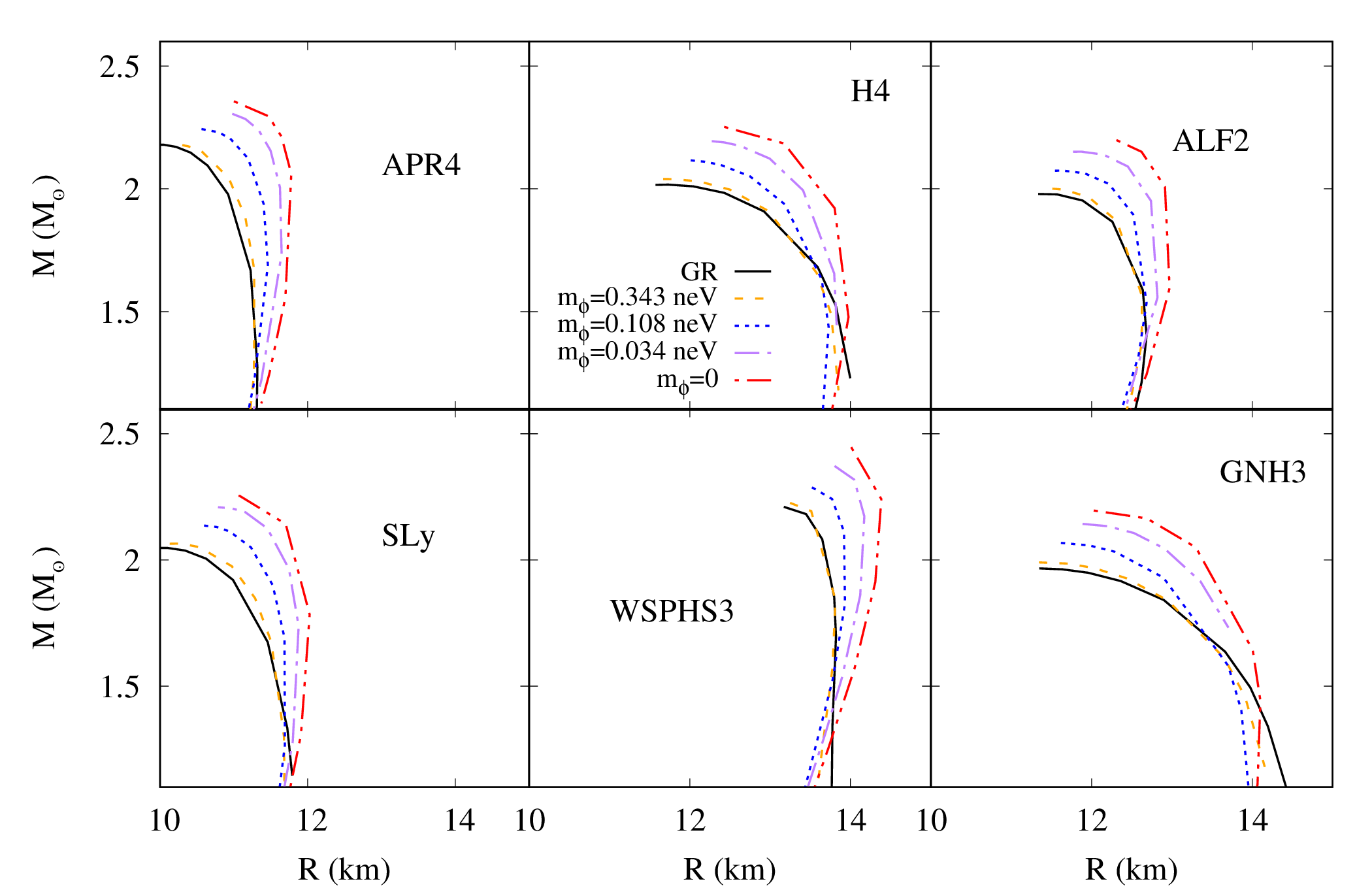}
	\includegraphics[width=.32\textwidth, angle =-90]{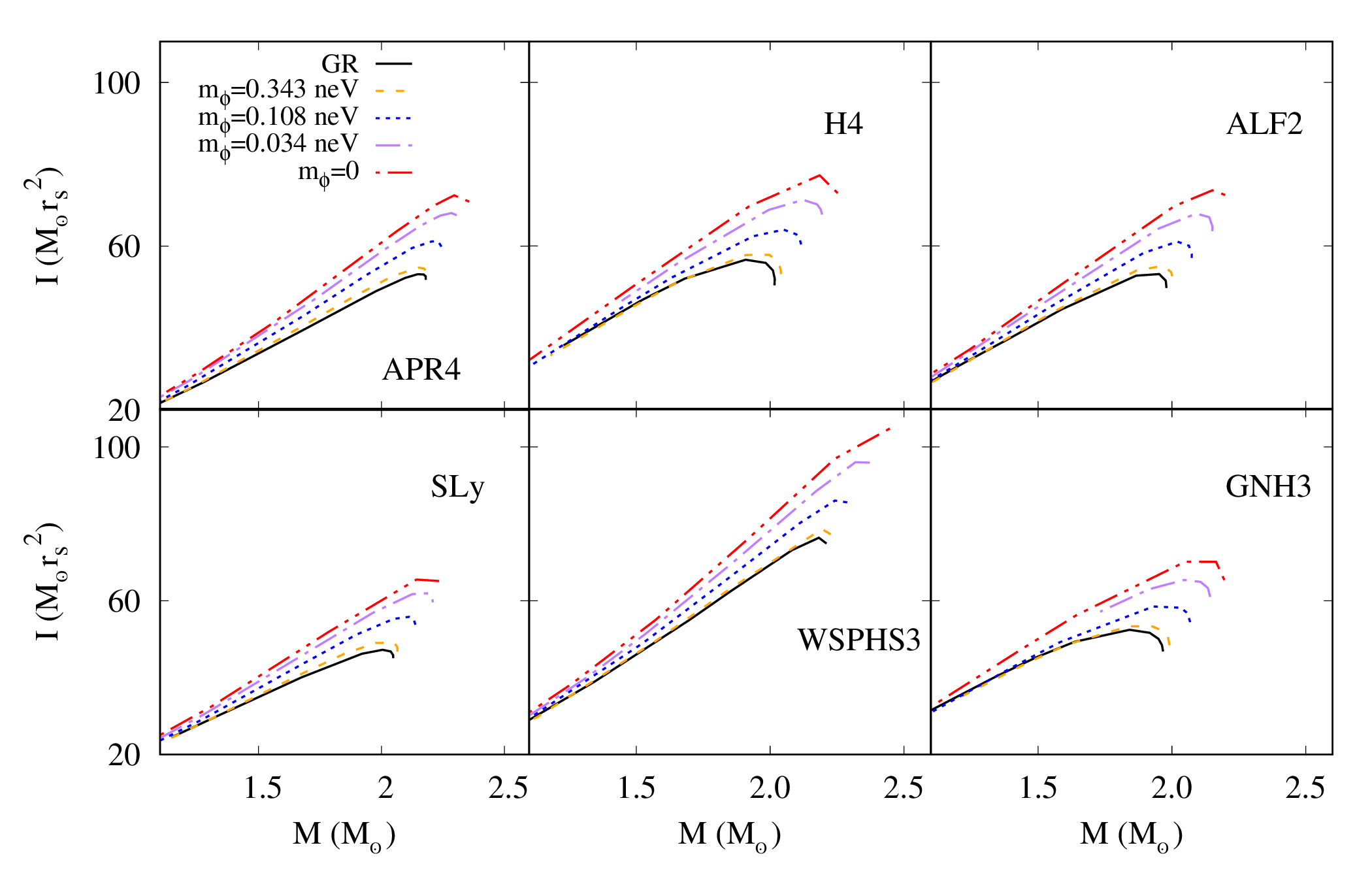}
	\caption{(left) Mass-radius {relation: mass $M$ in solar masses $M_\odot$ versus physical radius $R$ in km.} 
	(right) Moment of inertia-mass {relation: moment of inertia $I$ in units of $M_\odot r_S^2$ (with solar gravitational radius $r_s$). 
	The six panels represent six EOSs, and the colors indicate the values of the scalar field mass $m_{\phi}$, with the general relativistic limit in black.}
		 }
	\label{fig:M_I_panel_l1}
\end{figure}

For reference we include here {some relevant} results on the global properties of neutron stars. In Figure \ref{fig:M_I_panel_l1} (left) we show the mass-radius relation (see also \cite{Blazquez-Salcedo:2018qyy}).  In each panel we present a different EOS. Inside each panel, we show in black {the limiting case of} general relativity, in orange $m_\phi=0.343$ neV, in blue $m_\phi=0.108$ neV, in purple $m_\phi=0.034$ neV, and in red the massless limit. The radius $R$ (in km) is the physical radius of the star, i.e., the radius in the Jordan frame. {The mass $M$ (in solar masses $M_\odot$) is independent of the frame. The ratio of mass and radius yields the compactness $C=M/R$ of the star.}

Figure \ref{fig:M_I_panel_l1} (right) is an analogous figure for the moment of inertia $I$ as a function of the total mass $M$. The moment of inertia is given in units of $M_\odot r_s^2$, where $r_s=1.47664$ km is the {gravitational radius corresponding to one solar mass} $M_\odot$. Similar results have been obtained before in \cite{Staykov:2014mwa,Staykov:2015mma,Staykov:2016mbt}.
With the moment of inertia one can define the radius of gyration, $\hat{R}=\sqrt{I/M}$, and with this the so-called generalized compactness $\eta = M/\hat{R} = \sqrt{M^3/I}$. 

\subsection{Quadrupole f-mode: spectrum}

Here we present our results for the $l = 2$ perturbations. 
In this case, there are, apart from fluid-led modes and scalar-led modes, also curvature-led modes. 
{Previous investigation of the scalar-led $\phi$-modes revealed very little dependence of the frequency and the damping time on the mass of the neutron stars \cite{Blazquez-Salcedo:2021exm}.}
However, {the scalar-led and the curvature-led modes} are expected to be physically less relevant, since they possess shorter damping times (milliseconds), and they are expected to be more difficult to excite during astrophysical phenomena that produce gravitational waves. %

\begin{figure}[h!]
	\centering
	\includegraphics[width=.32\textwidth, angle =-90]{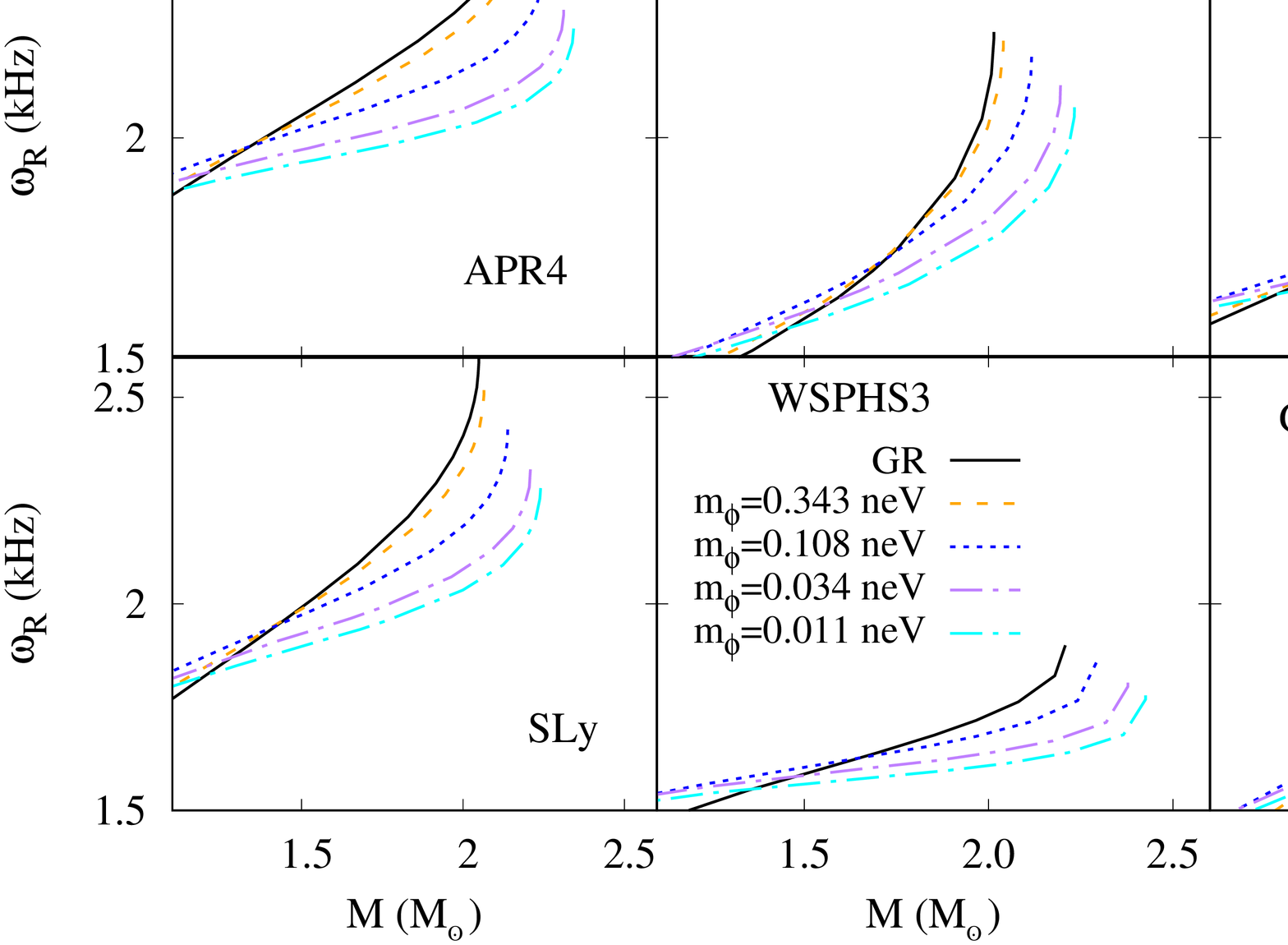}
		\includegraphics[width=.32\textwidth, angle =-90]{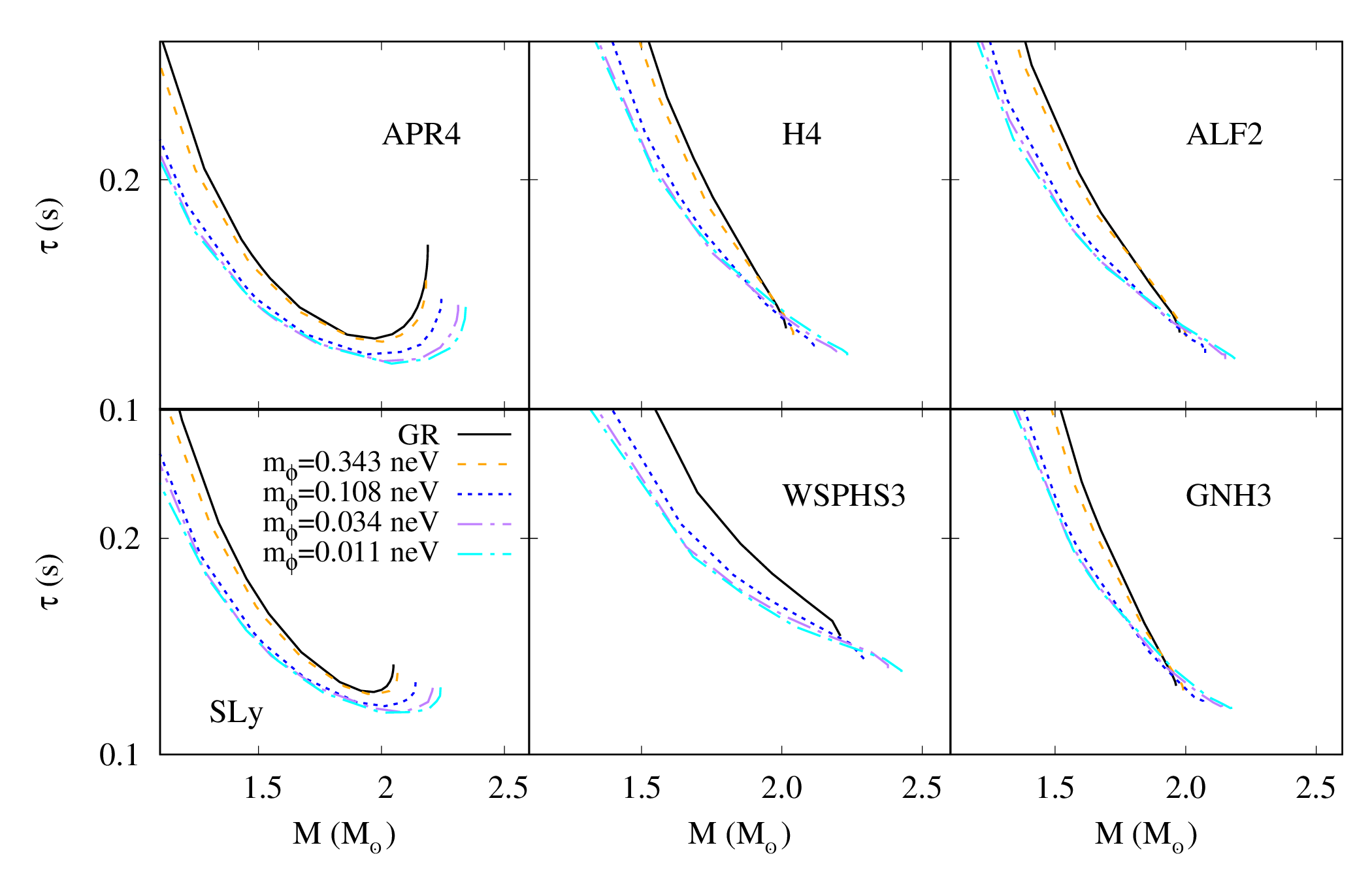}
	\caption{Frequency $\omega_R$ in kHz (left) and damping time $\tau$ in seconds (right) versus neutron star mass $M$ in $M_{\odot}$ for the fundamental quadrupole f-mode.
	{The six panels represent six EOSs, and the colors indicate the values of the scalar field mass $m_{\phi}$, with the general relativistic limit in black.}
		 }
	\label{fig:MR_MI_pcp_panel_l2}
\end{figure}

Hence, we here focus on the quadrupole f-mode. In Figure	\ref{fig:MR_MI_pcp_panel_l2} we show the frequency in kHz (left) and the damping time in seconds (right) for several EOSs (one per panel) and several values of the scalar mass (colored curves).
{For comparison, the corresponding f-modes of general relativity (black) are also shown}.
For heavier neutron stars, general relativity {leads} typically to the highest frequency values and the {longest} damping times. 
The effect of the scalar {degree of freedom} is to decrease these values, the lighter the scalar, the more important this effect. 
For small values of the neutron star mass, a crossing of the frequencies is observed, {such that} for intermediate values of the scalar mass, the frequencies may be slightly larger than the corresponding general relativistic values.

\subsection{Universal relations for the quadrupole f-mode}\label{Uni_rel_l2}

We now present our results concerning universal relations for the quadrupole f-mode.
We have performed our analysis for scalar field masses $m_\phi= 0.0108$ neV, 0.0343 neV, 0.1084 neV and 0.3428 neV and for the limiting case of general relativity including all six EOSs, SLy, APR4, H4, GNH3, ALF2 and WSPHS3.
Only for the highest scalar mass of $m_\phi=0.3428$ neV we have not included EOS WSPHS3, since the modes were very close to the modes of general relativity. 

The fitting procedure has been based on a polynomial of order 4, $f=ax^4+bx^3+ cx^2 + dx + e$.
When the fit error in a specific term went up to 100\% or beyond, the particular term has been considered redundant in the fitting procedure and has therefore been discarded.
In the following we exhibit a set of figures exploring various possibilities for universal relations for the frequencies $\omega_R$ and damping times $\tau$ of the modes, that involve a variety of properly scaled (in geometric units) dimensionless quantities on both axes.
For each fit we also present the corresponding fitting coefficients and errors in tables, 
{along with the average error of the fits for better comparison among the various relations}.
{For a better overview, the tables with the fit parameters for all these universal relations for the f-modes have been collected in Appendix \ref{Appendix_tables_l2}.}
We start our discussion with relations that involve only either the frequency $\omega_R$ or the damping time $\tau$, and continue subsequently with relations involving both $\omega_R$ and $\tau$.

\begin{figure}[H]
	\centering
	\includegraphics[width=.45\textwidth, angle =0]{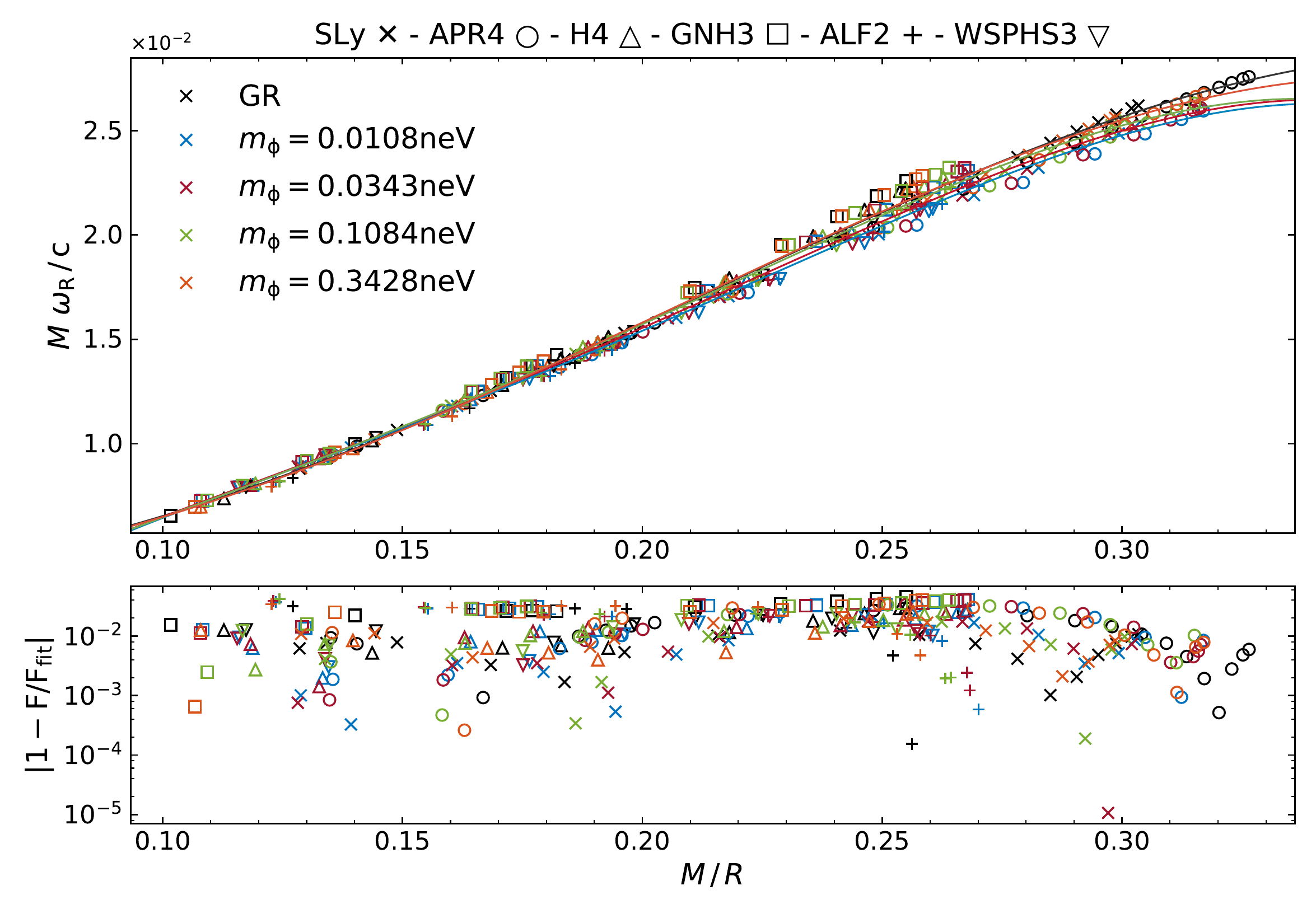}
	\includegraphics[width=.45\textwidth, angle =0]{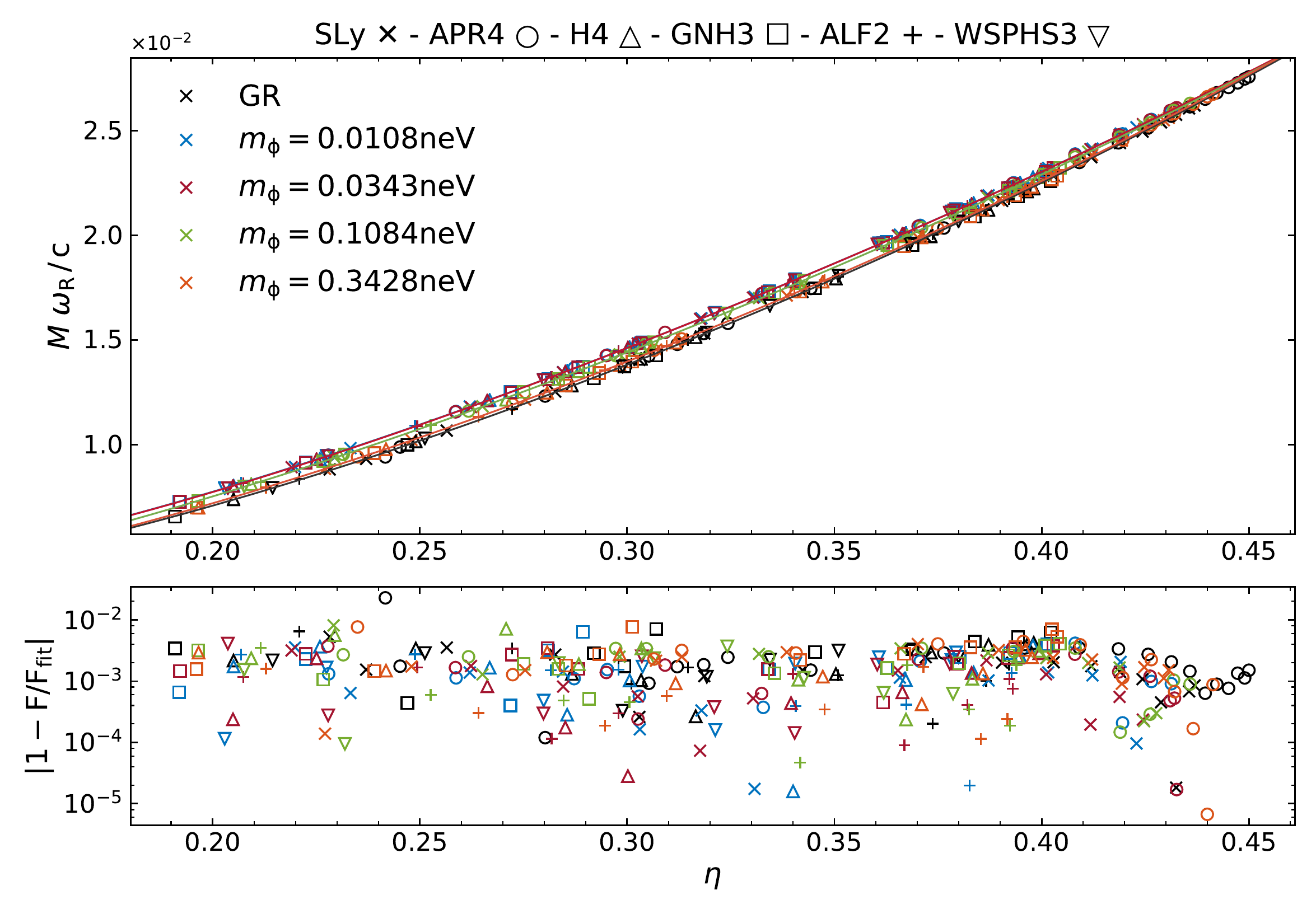}
	\caption{{f-mode universal relations: dimensionless frequency $M\omega_R/c$ (upper panels) and fit errors (lower panels) versus compactness $C=M/R$ (left panels); versus generalized compactness $\eta$ (right panels).
	The symbols indicate the respective EOS and the colors the values of the scalar field mass $m_{\phi}$ with the general relativistic limit in black.}
	}
	\label{fig:uni_rel_2_mass_omega}
\end{figure}

In Figure \ref{fig:uni_rel_2_mass_omega} we have scaled the frequency $\omega_R$ with the neutron star mass $M$, and exhibit the dimensionless frequency $M\omega_R/c$ versus the compactness $C=M/R$ on the left hand side of the figure in the upper panel. 
The various symbols indicate the numerical results of the calculations of the modes for their respective EOS.
At the same time the colors indicate the respective values of the scalar field mass $m_{\phi}$.
The general relativistic limit is shown in black.
For each theory, the curve with the respective color represents the corresponding best fit.
The lower panel contains the associated fit errors.
The right hand side is an analogous figure where the dimensionless frequency $M\omega_R/c$ is now shown versus the generalized compactness $\eta$ together with the fit errors.

The dimensionless f-mode frequency $M\omega_R/c$ exhibits a rather linear dependence on the compactness $C=M/R$ and (to a slightly lesser degree) on the generalized compactness $\eta$.
While a polynomial of degree 4 is used when fitting all results in this section to accommodate a larger class of dependencies between the dimensionless variables, for this particular example, almost all of the fit parameters of the higher-order terms ($a,b,c$) carry an error larger than 50\%, thus hinting at a linear fit.
We provide the corresponding values of the fit parameters and errors in Table \ref{rel_fit_MOmegaR_C_l2_lhs} and Table \ref{rel_fit_MOmegaR_eta_l2_rhs} of Appendix \ref{Appendix_tables_l2}. 
As noted already before in general relativity \cite{Lau:2009bu} and evident from Figure \ref{fig:uni_rel_2_mass_omega} also for $R^2$ gravity, the distribution of data points for the f-mode frequency exhibits a smaller fit error, when the generalized compactness $\eta$ is considered instead of the compactness $C$.

\begin{figure}[H]
	\centering
	\includegraphics[width=.45\textwidth, angle =0]{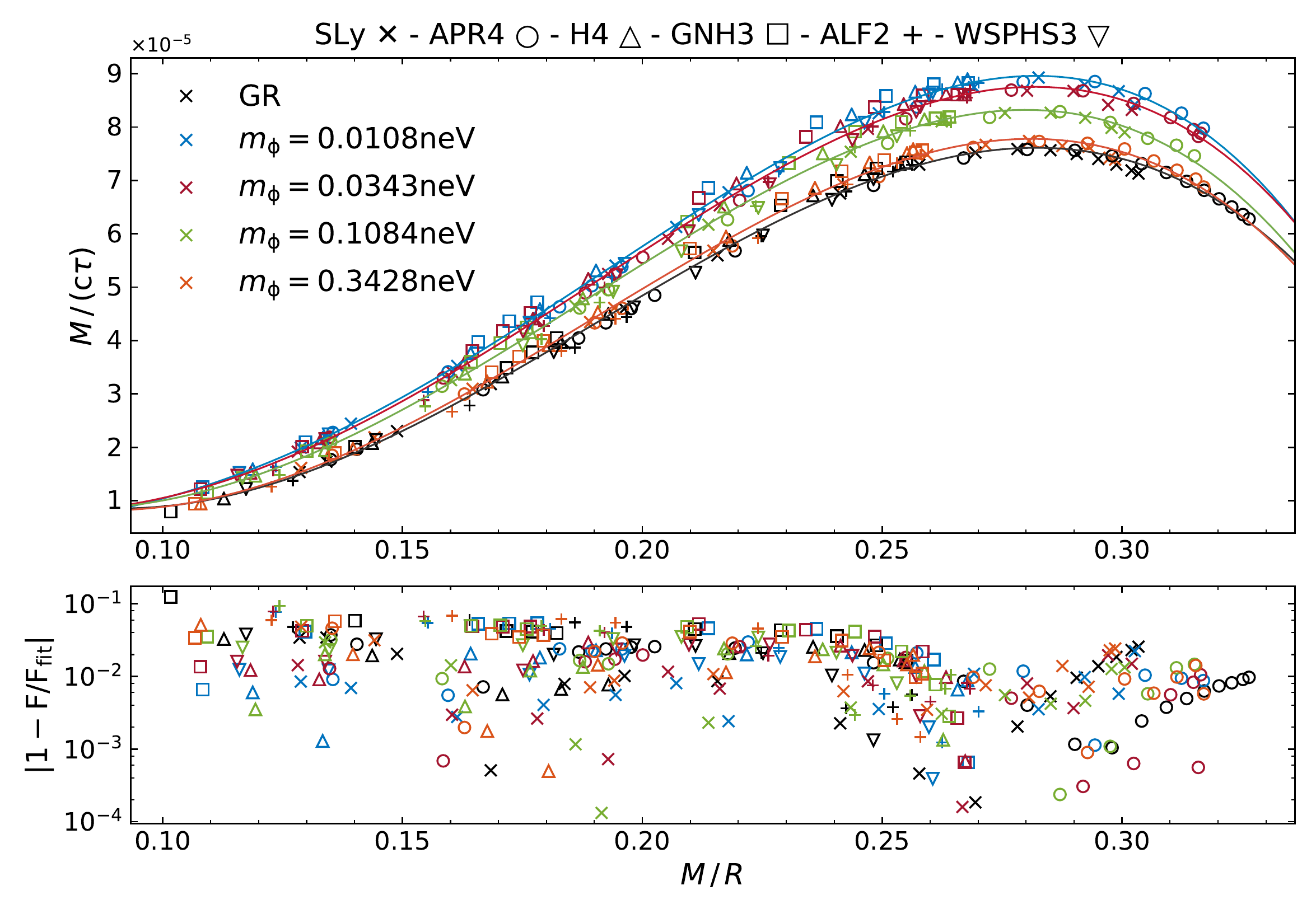}
	\includegraphics[width=.45\textwidth, angle =0]{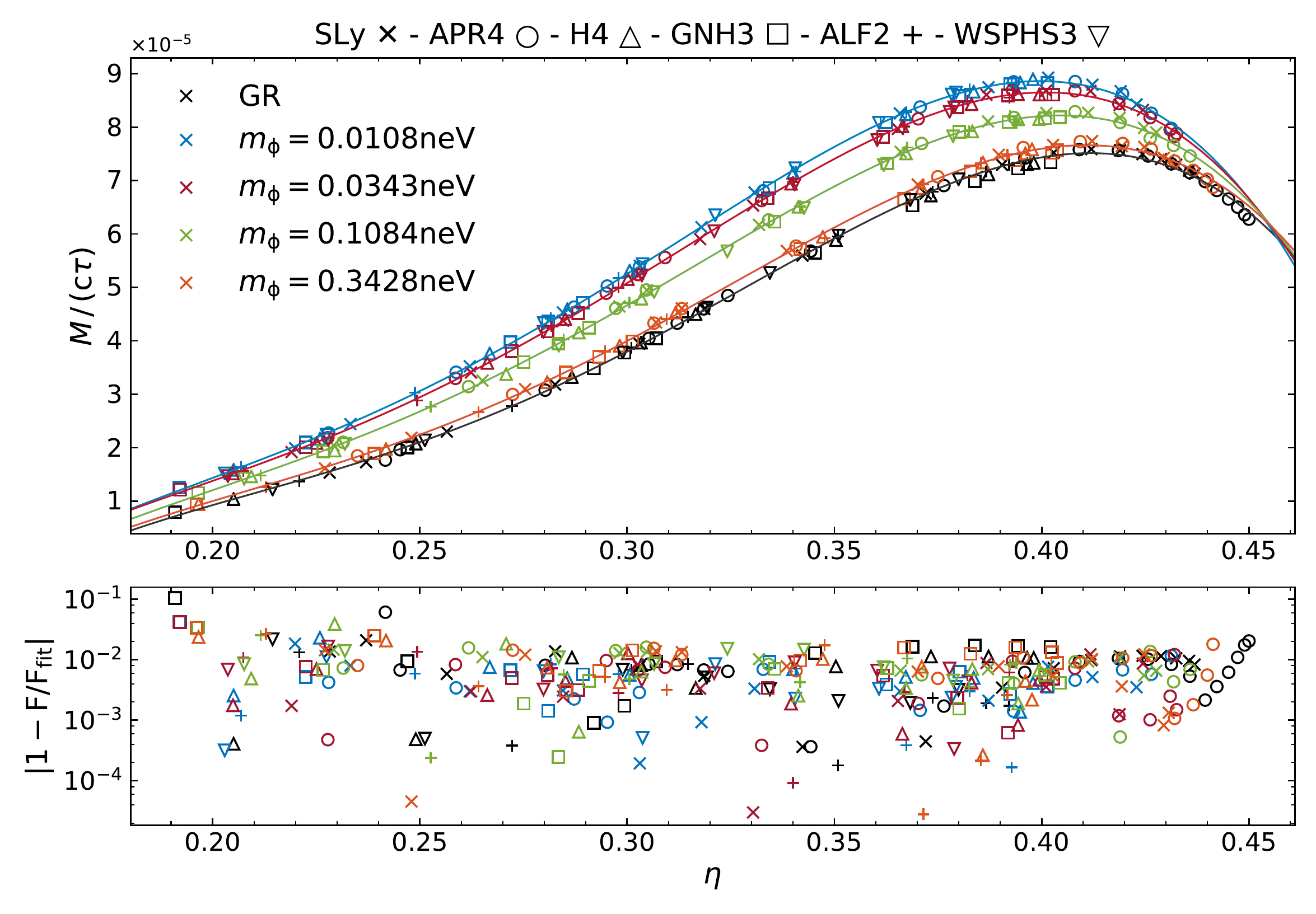}
	\caption{
	{f-mode universal relations: dimensionless inverse damping time $M/(c\tau)$ (upper panels) and fit errors (lower panels) versus compactness $C=M/R$ (left panels); versus generalized compactness $\eta$ (right panels).
	The symbols indicate the respective EOS and the colors the values of the scalar field mass $m_{\phi}$ with the general relativistic limit in black.}}
	\label{fig:uni_rel_mass_tau_full}
\end{figure}
Figure \ref{fig:uni_rel_mass_tau_full} is an analogous figure for the dimensionless inverse damping time $M/(c\tau)$, where the neutron star mass $M$ has also been used to obtain the relevant dimensionless quantity.
Again, the figure and also the corresponding tables for the fits, Tables \ref{rel_fit_MOmegaI_l2_lhs} and \ref{rel_fit_MOmegaI_eta_l2_rhs} of Appendix \ref{Appendix_tables_l2}, show that the errors in the fit parameters are generally lower when the generalized compactness $\eta$ is employed instead of the compactness $C$. {Comparing Figures \ref{fig:uni_rel_2_mass_omega} and \ref{fig:uni_rel_mass_tau_full} shows that the dimensionless frequency $M\omega_R/c$ provides a better fit than the dimensionless damping time $M/(c \tau)$. 
However, the dimensionless damping time $M/(c \tau)$ exhibits a larger split between the universal relations for the different scalar masses $m_\phi$.}
Such a split is of course needed, if one is to exploit the universal relations to obtain bounds on the theory.

\begin{figure}[H]
	\centering
	\includegraphics[width=.45\textwidth, angle =0]{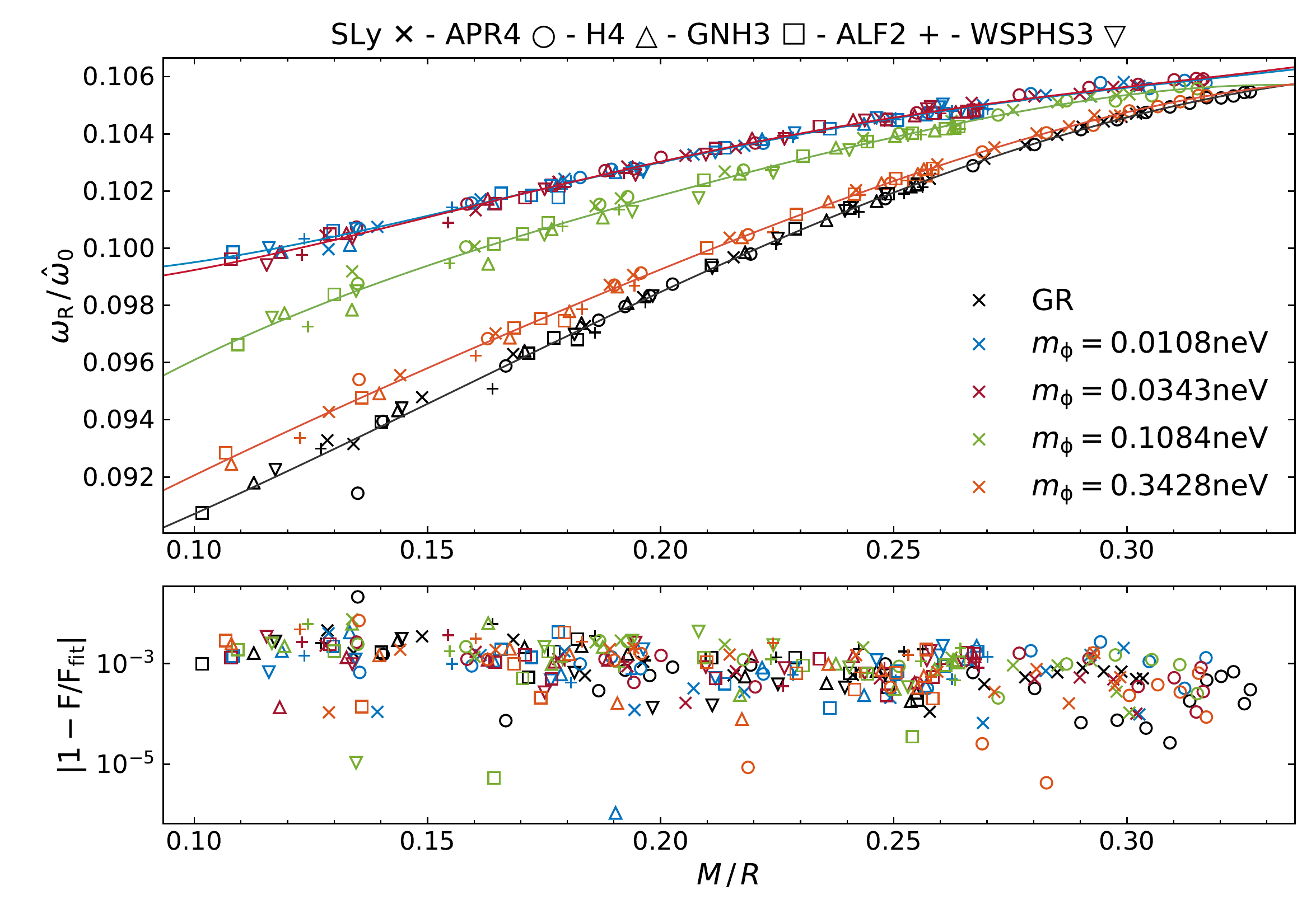}
	\includegraphics[width=.45\textwidth, angle =0]{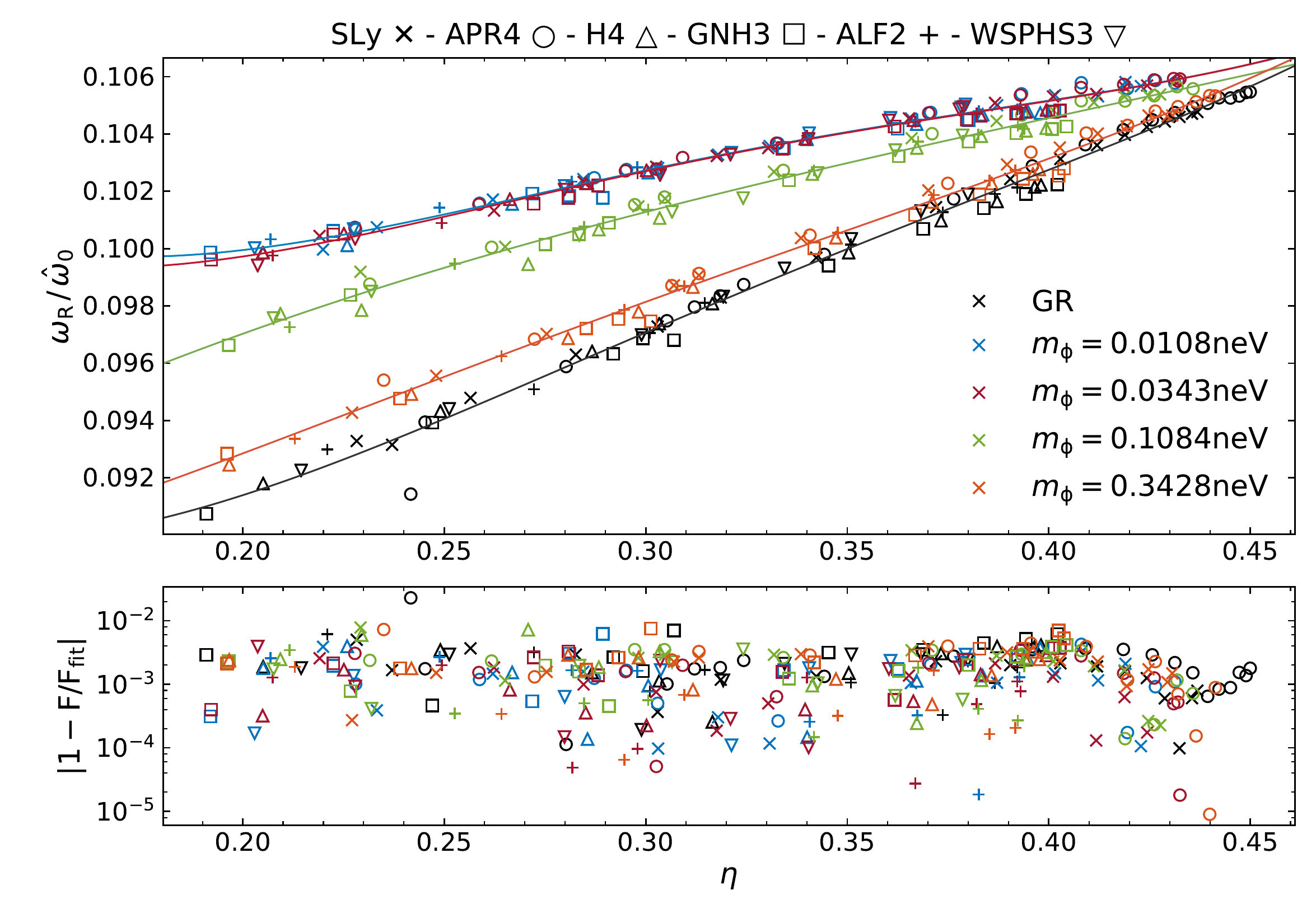}
	\caption{{f-mode universal relations: dimensionless frequency $\omega_R/\hat \omega_o$ 
	(upper panels) and fit errors (lower panels) versus compactness $C=M/R$ (left panels); versus generalized compactness $\eta$ (right panels).
	The symbols indicate the respective EOS and the colors the values of the scalar field mass $m_{\phi}$ with the general relativistic limit in black.}
	}
	\label{fig:uni_rel_omega_omegao}
\end{figure}

{Significant improvement of the fits is observed when we consider a different type of scaling of the frequency $\omega_R$, as shown in Figure \ref{fig:uni_rel_omega_omegao}.
Here we have employed the reference frequency $\hat \omega_o= \sqrt{\frac{3Mc^2}{4\hat{R}^3}}=\frac{c}{M}\sqrt{\frac{3}{4}\eta^3}$ (with $\hat{R}=\sqrt{I/M}$), that involves the generalized compactness. %
We exhibit the scaled frequency {$\omega_R/\hat{\omega}_o$} versus both compactness and generalized compactness.
While both are rather good, taking compactness instead of generalized compactness is {slightly} better.
Also here we observe the desired splitting of the universal relations, as the scalar mass is varied. {For the corresponding tables for the fits, see Table \ref{rel_fit_omega_scale_hat_l2_lhs} and \ref{rel_fit_omega_scale_hat_l2_rhs}.}
We have also performed the analysis with the reference frequency $\omega_o= \sqrt{\frac{3Mc^2}{4{R}^3}}=\frac{c}{M}\sqrt{\frac{3}{4}C^3}$, that is analogous but involves compactness \cite{Lindblom:1989}.
The corresponding fits fare much worse again, therefore we do not show them here.}
{
{Apart from that}, the dimensionless frequency $\omega_R/\hat{\omega}_o$ 
{also}
yields a much better universal relation
than the dimensionless quantities $R\omega_R/(c\eta)$ {and} $R\omega_R/(c\eta^3)$,
{with the latter being the worst of all}.
}

\begin{figure}[H]
	\centering
	\includegraphics[width=.45\textwidth, angle =0]{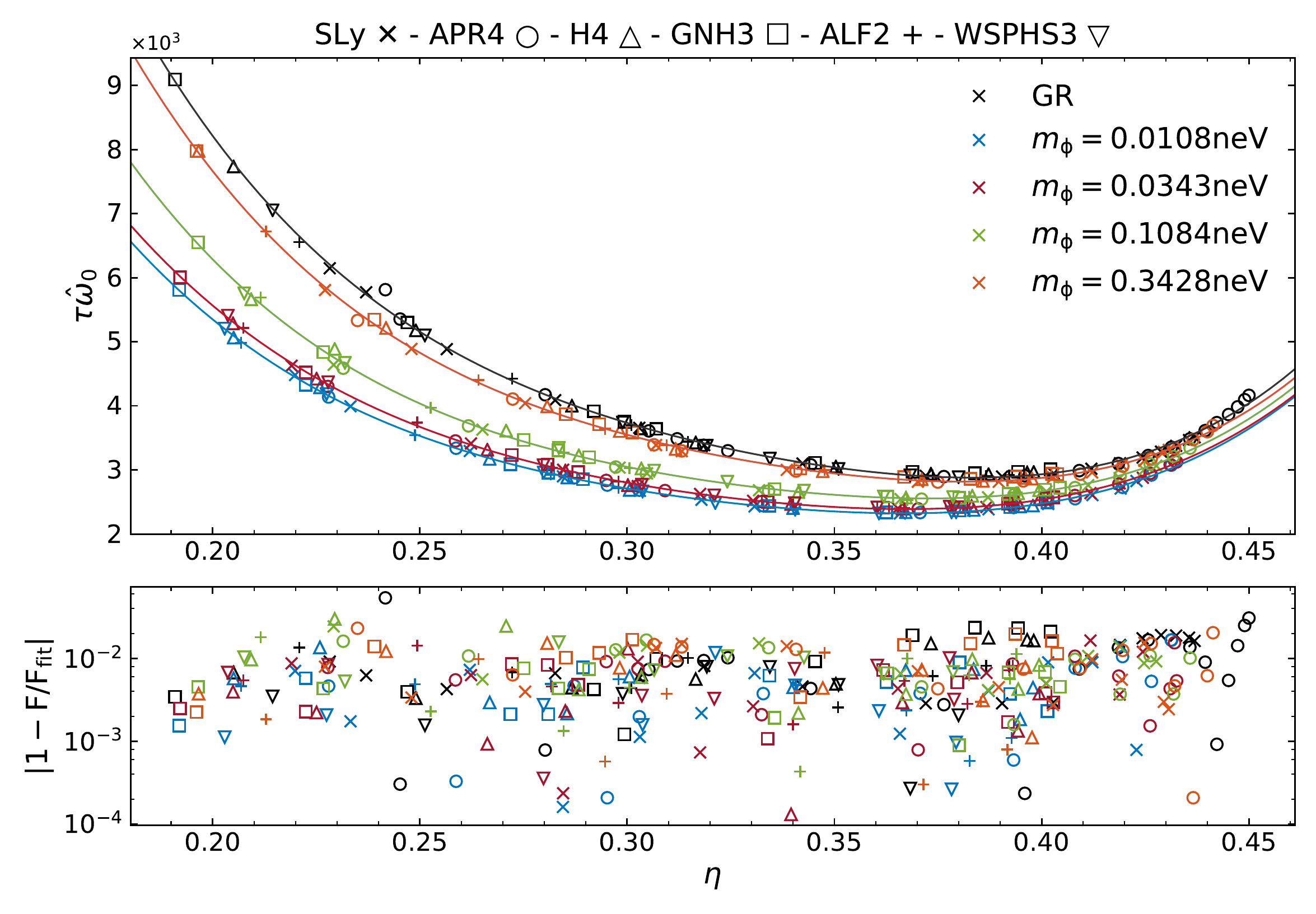}
	\includegraphics[width=.45\textwidth, angle =0]{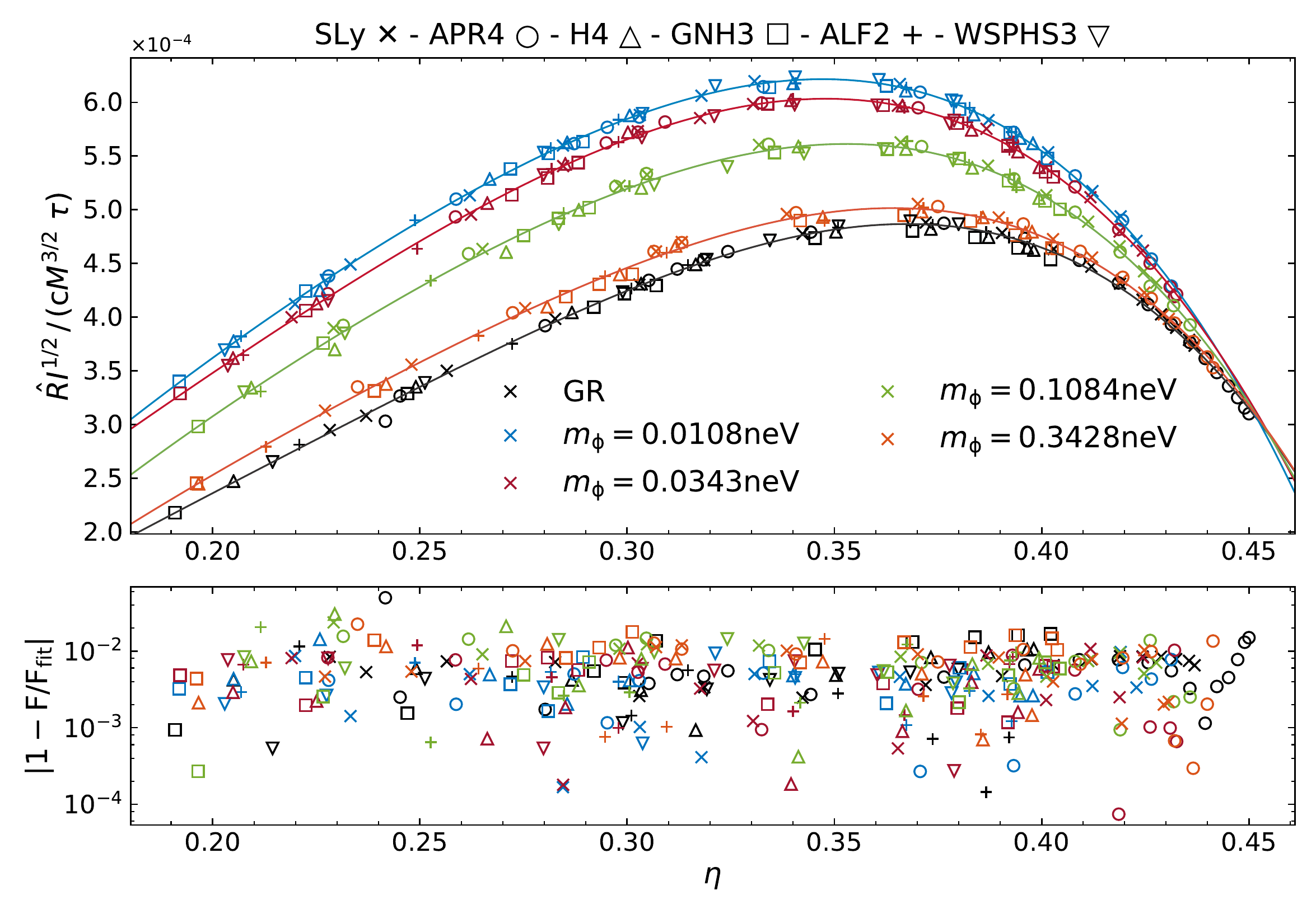}
	\caption{{f-mode universal relations: dimensionless damping time $\tau \hat \omega_o$ 
	versus generalized compactness $\eta$
	(left upper panel)
	and fit errors (left lower panel); 
	dimensionless function of damping time
	$\hat{R}I^{1/2}/(cM^{3/2}\tau)$
	versus generalized compactness $\eta$ (right upper panel) and fit errors (right lower panel).
	The symbols indicate the respective EOS and the colors the values of the scalar field mass $m_{\phi}$ with the general relativistic limit in black.}
	}
	\label{fig:uni_rel_tau_omegao}
\end{figure}

{ 
Figure \ref{fig:uni_rel_tau_omegao} (left) exhibits the fits for the damping time scaled with $\hat \omega_o$. Here we consider the scaled damping time {$\tau\hat{\omega}_o$} versus the generalized compactness. 
Table \ref{rel_fit_tau_scale_hat_l2_rhs} shows the corresponding fit parameters.
{As opposed to} the scaled frequency, we observe an improvement for the scaled damping time, with the generalized compactness faring slightly better than the compactness.
Again, when employing the reference frequency $\omega_o$ we see worse fits, so we neither display them here.
We have also investigated the scaled frequency $R\omega_R/(c\eta^3)$ and the scaled damping time $R/( c\eta^3 \tau)$, but the results of the fits were again worse than those above based on scaling with $\hat \omega_o$.
Figure \ref{fig:uni_rel_tau_omegao} (right)
shows the relations involving the damping time $\hat{R} I^{1/2}/ (c M^{3/2} \tau)$ and the generalized compactness,
with improved errors,
where the fit parameters can be found in 
Table \ref{rel_fit_tau_scale_hat_l2_lhs}.}

{ 
The universal relations with the dimensionless function of the damping time 
$R^4/ (c M^3 \tau)=R/(c C^3 \tau)$ and the compactness $C$
possess an error bounded by at most 10\%. 
This fit is worse as the mean errors from the theories are considerably larger than those produced by the relations in Figure \ref{fig:uni_rel_tau_omegao} (right).
{It does not improve if one replaces $C$ by $\eta$, i.e. $R/(c\eta^3\tau)$.}
However, if one considers $R/(c\eta^3\tau)$ against the generalized compactness, the average error {improves slightly}.
{We also tested the universal relation of $R/(c\eta\tau)$ with the compactness $C$ and the generalized compactness $\eta$. The {average} error of these fits is roughly the same as when scaling the damping time with $\hat{\omega}_o$, 
{although it results in a clearer splitting of the universal relations for different scalar masses}.}
We find that the average errors can be even lowered when we 
consider universal relations utilizing the radius of gyration $\hat{R}$ in a dimensionless function of the real frequency
as in $\hat{R}\omega_R I^{1/2}/ (c M^{3/2})$ and the generalized compactness $\eta$, 
shown in Figure \ref{fig:uni_rel_mass_tau__mass_omega_full} (left).
Table \ref{rel_fit_R4_l2_rhs} of Appendix \ref{Appendix_tables_l2} presents the associated fits.
They give an average error slightly smaller than their imaginary frequency counterpart in Figure \ref{fig:uni_rel_tau_omegao} (right). 
\begin{figure}[H]
	\centering
	\includegraphics[width=.45\textwidth, angle =0]{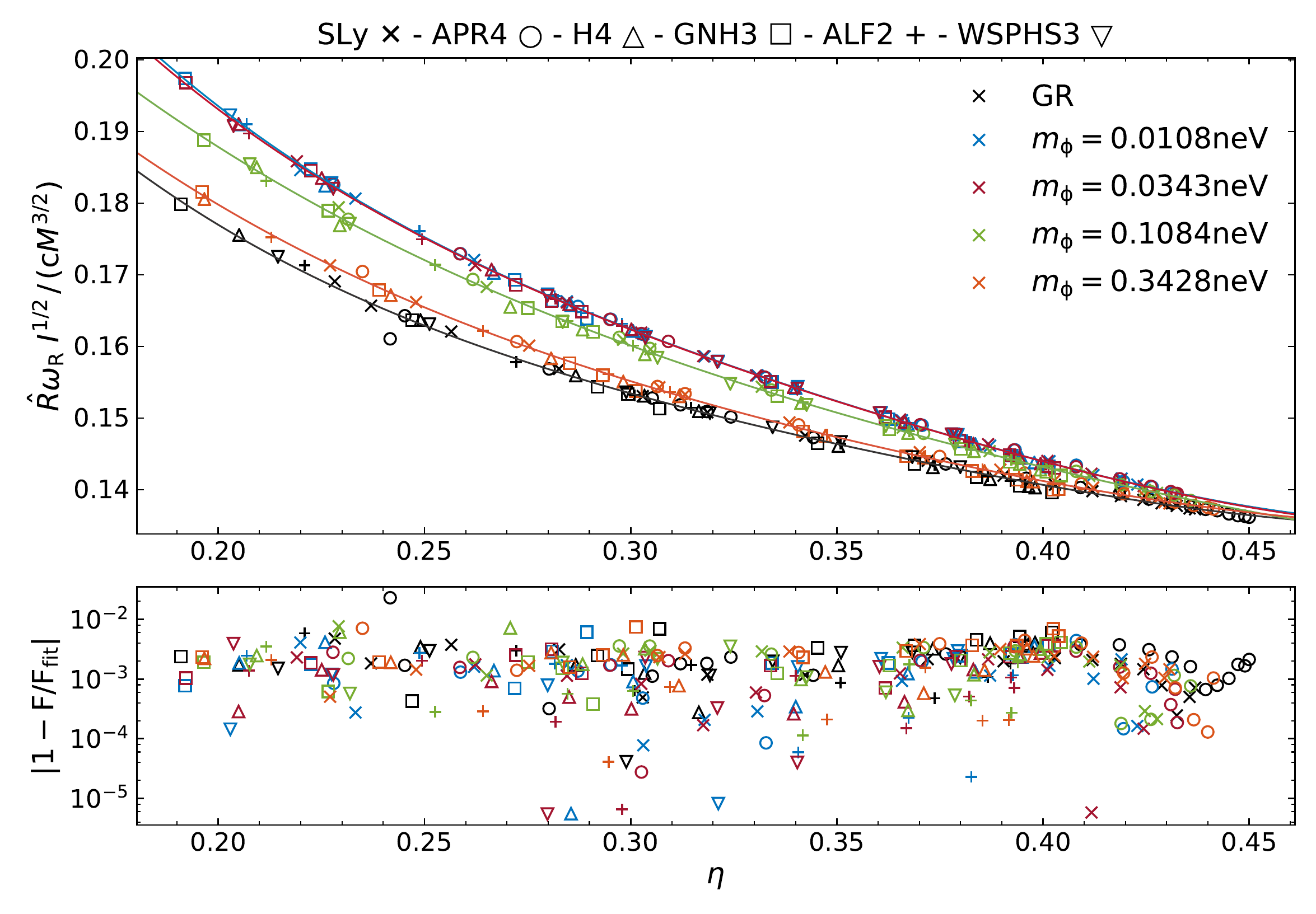}
	\includegraphics[width=.45\textwidth, angle =0]{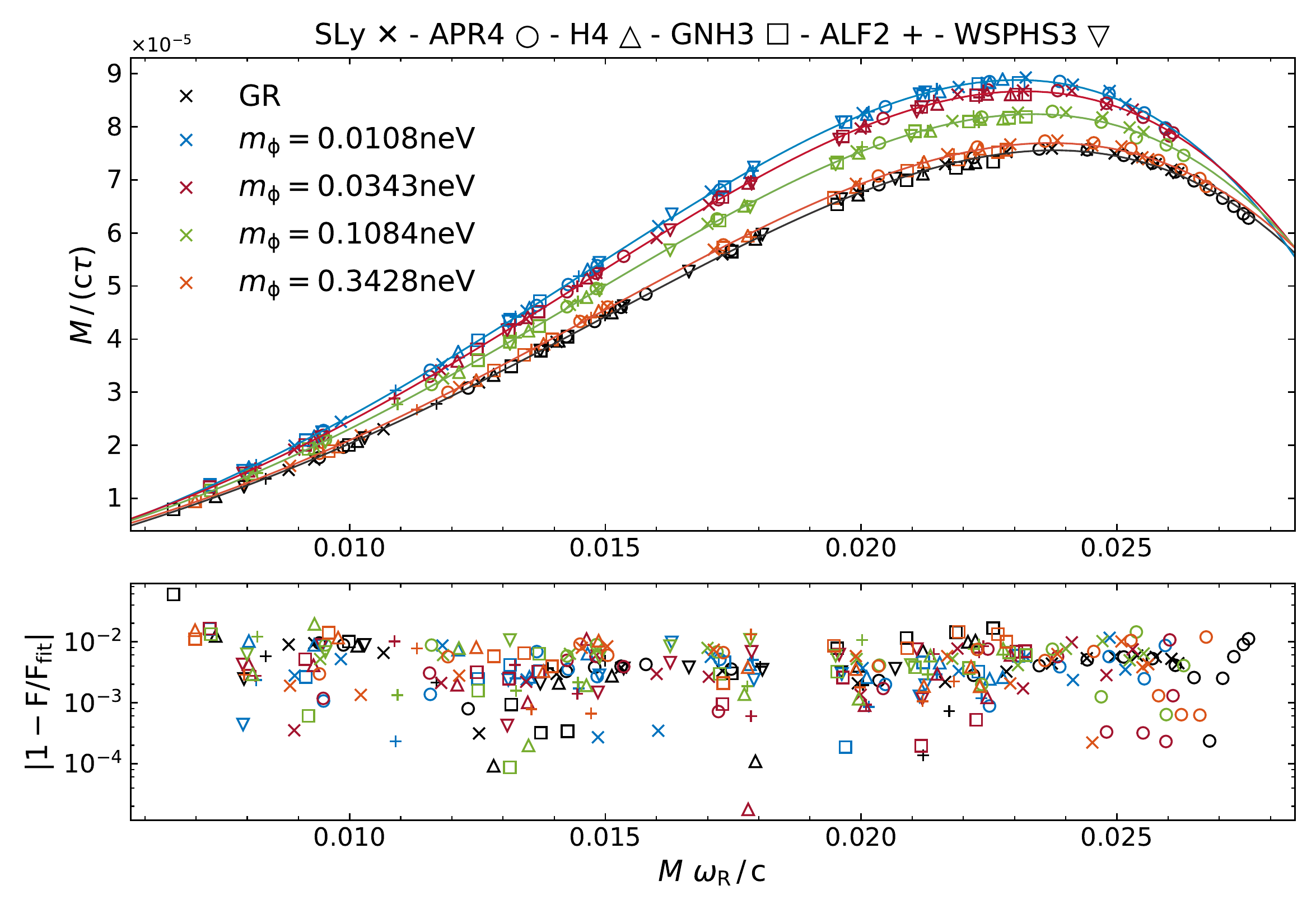}
	\caption{
	{f-mode universal relations: dimensionless function of real frequency $\hat{R}\omega_R I^{1/2}/ (c M^{3/2})$ versus compactness $\eta=M/\hat{R}$ (left upper panel) and fit errors (left lower panel);
	dimensionless inverse damping time $M/(c\tau)$ versus dimensionless frequency $M\omega_R/c$ (right upper panel) and fit errors (right lower panel). 
	The symbols indicate the respective EOS and the colors the values of the scalar field mass $m_{\phi}$ with the general relativistic limit in black.}
	}
	\label{fig:uni_rel_mass_tau__mass_omega_full}
\end{figure}
}

We now consider universal relations containing both the frequency $\omega_R$ and the damping time $\tau$.
In Figure \ref{fig:uni_rel_mass_tau__mass_omega_full} (right) we display universal relations where the dimensionless inverse damping time $M/(c\tau)$ is considered as a function of the dimensionless frequency $M\omega_R/c$, and we have scaled with the neutron star mass $M$.
The associated fits are presented in Table \ref{rel_fit_tau-OmegaR_l2_lhs} of Appendix \ref{Appendix_tables_l2}.
These universal relations turn out to be currently the best fits that involve both $\omega_R$ and $\tau$.

\begin{figure}[H]
	\centering
	\includegraphics[width=.45\textwidth, angle =0]{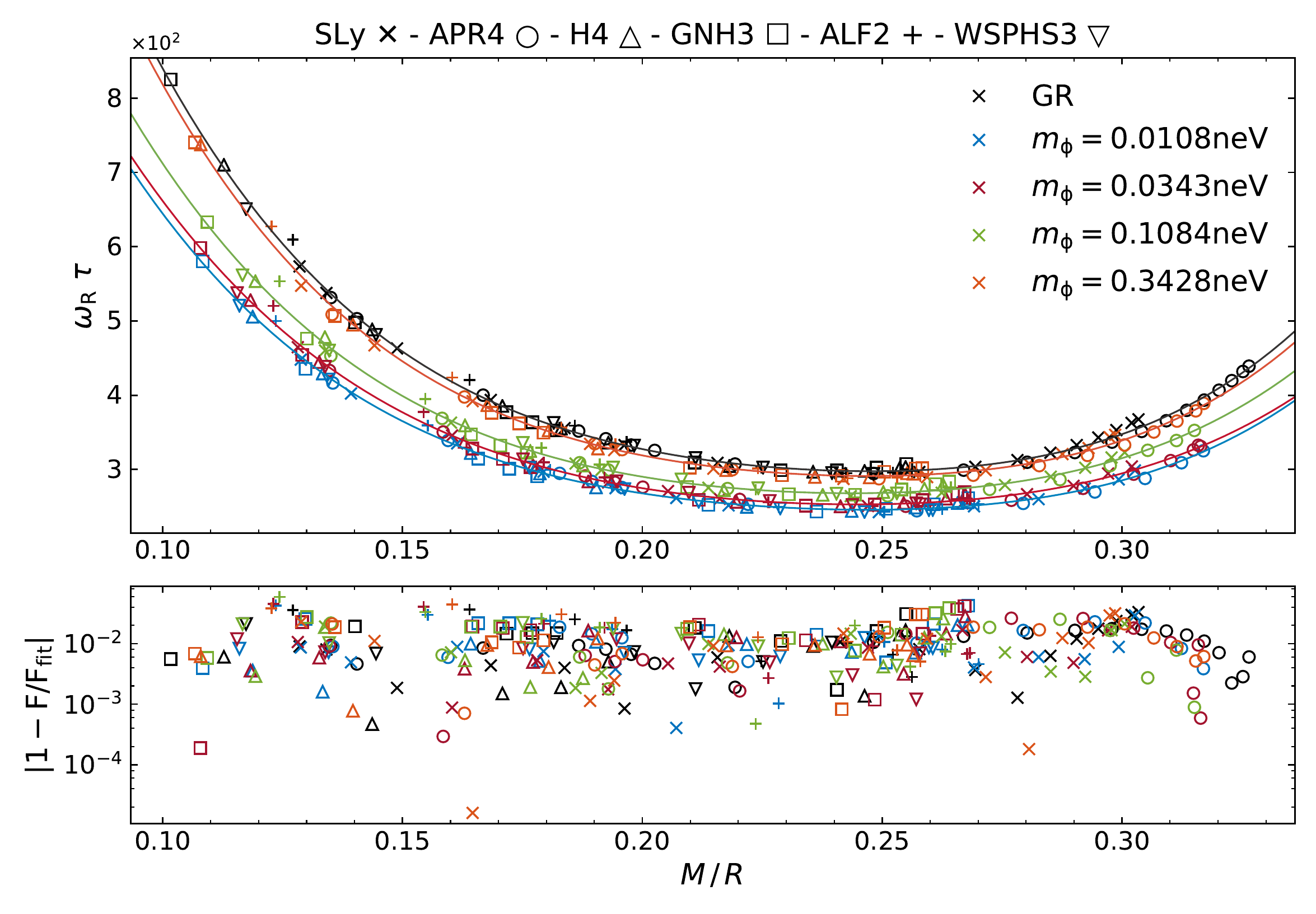}
	\includegraphics[width=.45\textwidth, angle =0]{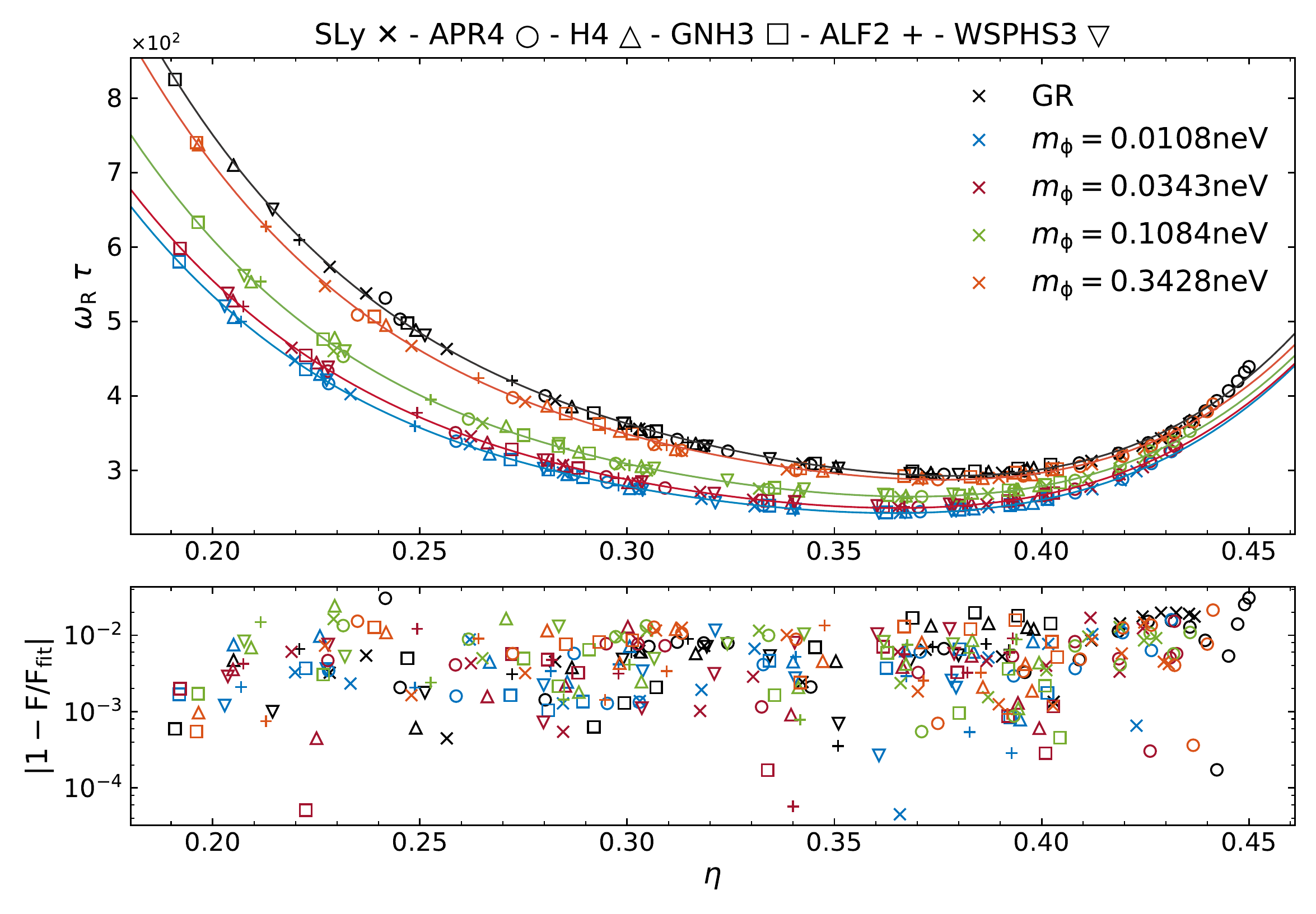}
	\caption{
	{f-mode universal relations: dimensionless product of frequency and damping time $\omega_R\, \tau$ (upper panels) and fit errors (lower panels) versus compactness $C=M/R$ (left panels); versus generalized compactness $\eta$ (right panels).
	The symbols indicate the respective EOS and the colors the values of the scalar field mass $m_{\phi}$ with the general relativistic limit in black.}
	}
	\label{fig:uni_rel_omega_tau_full}
\end{figure}
For comparison we also exhibit universal relations containing the product of the frequency $\omega_R$ and the damping time $\tau$.
Figure \ref{fig:uni_rel_omega_tau_full} exhibits this product $\omega_R\, \tau$ versus compactness $C$ (left) and generalized compactness $\eta$ (right).
The respective fit parameters are shown in Table \ref{rel_fit_tauOmegaR_l2_lhs} and Table \ref{rel_fit_tauOmegaR_eta_l2_rhs} of Appendix \ref{Appendix_tables_l2}.
The quality of the fits is reasonable but represents no improvement, quite to the contrary.
{{Other universal relations we tested, such as 
$M \tau \omega_R^2 /c$ 
and 
$R \tau \omega_R^2 /c$ 
versus $C$ and $\eta$, are not convincing either and we refrain from exhibiting them here.}}

{  
Among the universal relations which contain solely the 
frequency, the relations of
$\omega_R/\hat{\omega}_o$ versus the compactness are the best, while
for damping time (e.g. in Figure \ref{fig:uni_rel_mass_tau_full} and Figure \ref{fig:uni_rel_tau_omegao} (left)),
the universal relations of 
$\hat{R} I^{1/2}/ (c M^{3/2} \tau)$ and generalized compactness
are the best.
In the case when both frequency and damping time are involved, 
$\hat{R} I^{1/2}/ (c M^{3/2} \tau)$ relations are comparable to the relations of $\omega_R\tau$,
and the best relations are given by $M/(c\tau)$ and $M\omega_R/c$. 
}

{Concluding we note, that the f-mode can be rather well described in terms of universal relations involving scaling with $\hat \omega_o$, shown in Figure \ref{fig:uni_rel_omega_omegao} and \ref{fig:uni_rel_tau_omegao}, 
{as is as well evident from their lower average error $\bar{\epsilon}$}.
Other relations tested range from reasonable to rather unconvincing.
Most of the universal relations reveal a monotonic change with the scalar mass $m_\phi$, smoothly reaching the general relativistic limit for sufficiently large scalar mass.
Depending on the accuracy of future observations of the gravitational radiation from neutron stars, the split might suffice to put stronger lower bounds on the coupling constant $a$ and the associated scalar mass $m_\phi$.}


\subsection{Dipole F-mode: spectrum}\label{Dipole_spectrum}

We now turn to the dipole ($l=1$) perturbations. 
Analogous to the radial ($l=0$) sector, the dipole perturbations possess fluid-led modes and scalar-led modes. Here we will focus on the fundamental fluid mode, the dipole F-mode. 
Dipole modes were studied previously for several EOSs in general relativity (see e.g. \cite{Lindblom:1989,Takata:2008}).
The most important point here is, however, that dipole modes (like radial modes) do not propagate in general relativity.
Therefore the modes are normal modes, that possess real eigenvalues, i.e, only a frequency $\omega_R$, while $\omega_I=0$.
Consequently, when the general relativistic modes are smoothly reached in the limit of infinite scalar mass $m_\phi$, the imaginary part of the mode eigenvalue should decrease to zero,
making the modes ultra-long lived, as long as $m_\phi$ is finite, and approach the non-propagating modes of general relativity in the limit.

\begin{figure}[H]
	\centering
	\includegraphics[width=.32\textwidth, angle =-90]{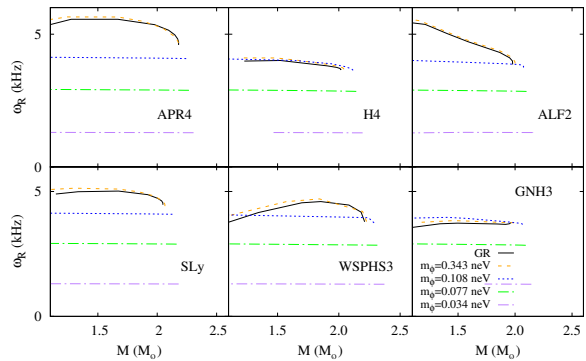}
	\caption{
	Frequency $\omega_R$ in kHz versus neutron star mass $M$ in $M_{\odot}$ for the fundamental dipole F-mode.
	{The six panels represent six EOSs, and the colors indicate the values of the scalar field mass $m_{\phi}$, with the general relativistic limit in black.}
		 }
	\label{fig:M_OmegaR_pcp_panel_l1}
\end{figure}
In Figure \ref{fig:M_OmegaR_pcp_panel_l1} we show the frequency $\omega_R$ of the dipole F-mode (in kHz) versus the mass $M$ (in $M_\odot$) analogously to the quadrupole f-mode in Figure	\ref{fig:MR_MI_pcp_panel_l2}, i.e., the panels correspond to different EOSs, and the colors to different values of the scalar field mass $m_\phi$ with general relativity in black.
{The figure contains only the real part of the mode eigenvalue and not the imaginary part, however, since the $\omega_I$ is very small and therefore cannot be extracted with sufficient accuracy.
The smallness of $\omega_I$ then translates into the largeness of its inverse $\tau$, making these dipole F-modes ultra-long lived.
At the same time, this conforms to the above expectation of approaching the general relativistic limit of non-propagating modes.}

{When recalling and comparing with the pattern observed for the radial F-mode in $R^2$ gravity \cite{Blazquez-Salcedo:2020ibb,Blazquez-Salcedo:2021exm} we see a strong correspondence.
The frequencies $\omega_R$ are %
almost independent of the neutron star mass, except for the scalar masses closest to general relativity and neutron star masses close to the maximum mass.
Moreover, both the radial F-mode and the dipole F-mode are ultra-long lived in the full range of scalar masses studied.
A clear difference is, however, their behavior along a family of neutron stars rather close to the maximum neutron star mass.
The radial F-mode frequency must decrease to zero, since the radial mode signals the instability of neutron stars beyond the maximum mass.
There the eigenvalue of the mode becomes purely imaginary, indicating the instability timescale.
Since the dipole F-mode frequency does not decrease to zero towards the maximum mass, this signals that this mode does not carry a further instability.}

{When considering the frequency $\omega_R$ as a function of the scalar mass $m_\phi$, the analysis of the radial F-mode revealed, that its scale is set by the size of the neutron star as long as the Compton wavelength  {$L_\phi$} corresponding to $m_\phi$ is small with respect to the size.
However, when the Compton wavelength is large as compared to the size of the neutron star, it is simply the scalar mass $m_\phi$ itself that sets the scale of the frequency.
Moreover, in this case this behavior of the scalar F-mode follows roughly the scalar $\phi$-mode \cite{Blazquez-Salcedo:2020ibb,Blazquez-Salcedo:2021exm}.
Since the dipole F-mode shows so much similarity to the radial F-mode, a similar pattern may be present also in this respect.
While this is supported by our current data, a systematic study of the dependence on the scalar mass $m_\phi$ has not yet been made.
}

\begin{figure}[H]
	\centering
	\includegraphics[width=.45\textwidth, angle =0]{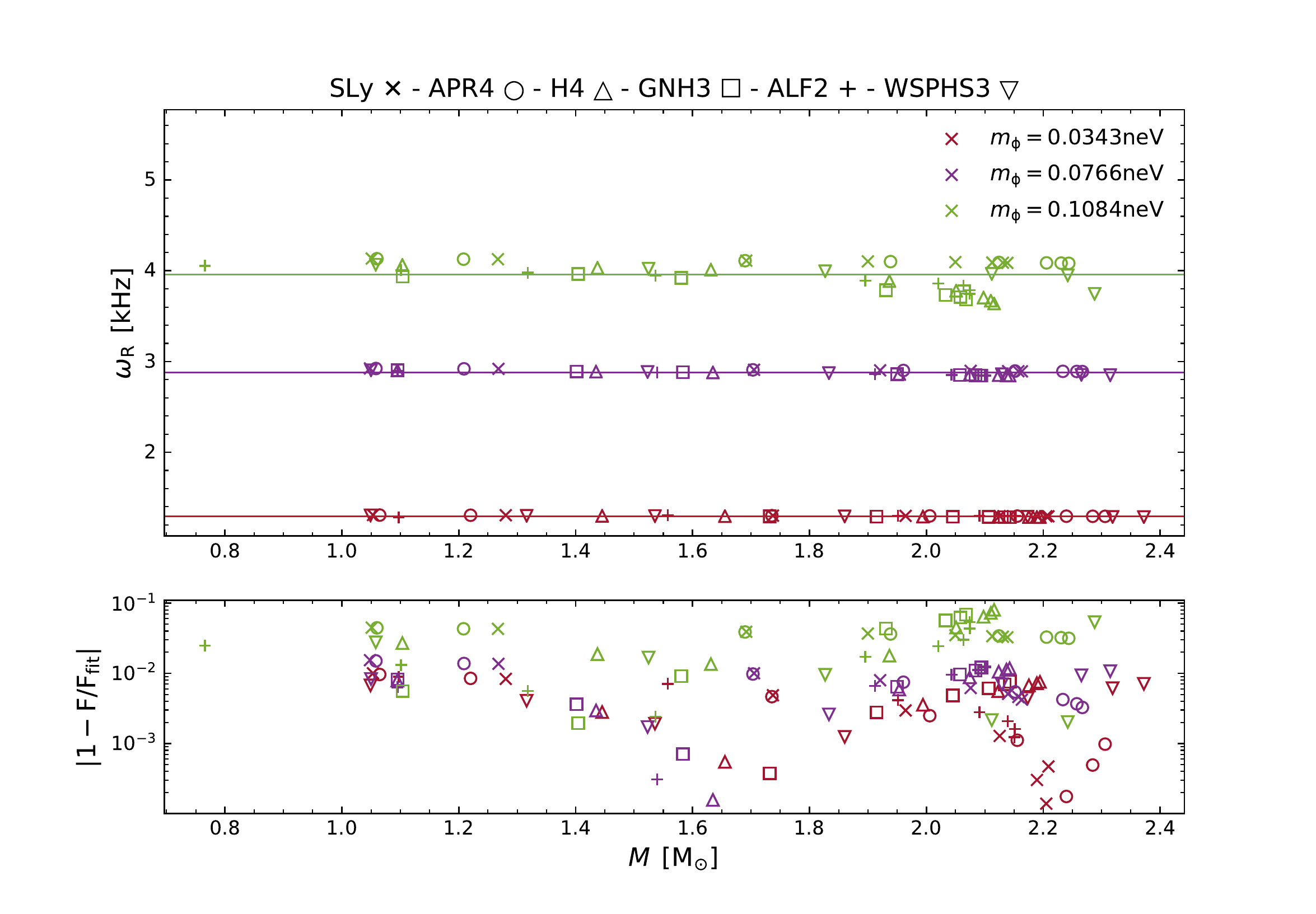}
	\caption{
	{F-mode analysis: frequency $\omega_R$ versus mass $M$ (upper panel) and fit errors (lower panel) for a constant fit. %
	The symbols indicate the respective EOS and the colors the values of the scalar field mass $m_{\phi}$ for $m_{\phi}=0.108$ neV,  $0.0767$ neV, and $0.0343$ neV.}
	}
	\label{fig:analysis_l1}
\end{figure}

{However, our current study lends additional support to this conjecture that arises from the investigation of the mode for a set of EOSs.
Here Figure \ref{fig:M_OmegaR_pcp_panel_l1} shows, that the value of the frequency is not only rather independent of the neutron star mass but also of the employed EOS.
An analysis of this independence is shown in Figure \ref{fig:analysis_l1} {for the lower values of the scalar field mass}, where we exhibit the frequency $\omega_R$ versus the neutron star mass $M$ for all EOSs, as if it where a universal relation, with the corresponding constant fit and error analysis. %
{The corresponding fit parameters including the errors can be found in Table \ref{rel_fit_analysis_l1_lhs} 
in Appendix \ref{App_dipole}.}
The constant fit yields for $m_{\phi}=0.108$ neV a frequency $\omega_R =3.96$ kHz, for $m_{\phi}=0.0767$ neV: $\omega_R =2.88$ kHz, and for $m_{\phi}=0.0343$ neV: $\omega_R =1.295$ kHz. 
Note, that the respective values of the frequency would be $\omega_R =4.16$ kHz for $m_{\phi}=0.108$ neV, $\omega_R =2.95$ kHz for $m_{\phi}=0.0767$ neV, and $\omega_R =1.32$ kHz for $m_{\phi}=0.0343$ neV, if we simply employed the formula $2\pi \omega_R=c/L_\phi$ found for the $l=0$ F-mode.
Here this behavior is seen up to somewhat higher scalar masses, and is ending when general relativity dominates the behavior of the $l=1$ F-modes.
Clearly, for the two lowest scalar masses $m_{\phi}=0.0343$ neV and $0.0767$ neV the constant fit is rather convincing, while the quality declines for $m_{\phi}=0.108$ neV as can be expected towards larger scalar mass.
}

{The above observation makes in principle a search for universal relations for sufficiently small scalar masses in $R^2$ theory superfluous.
It would fully suffice to %
measure the frequency $\omega_R$ of the F-mode to determine the scalar field mass and thus the theory.
It would be interesting though to investigate, whether other theories involving scalars would exhibit an analogous phenomenon, and if they would, to see whether there would be a way to distinguish among the theories.
}

\subsection{Universal relations for the ultra-long lived dipole F-modes}\label{Uni_rel_l1}

\begin{figure}[h!]
	\centering
		\includegraphics[width=.45\textwidth, angle =0]{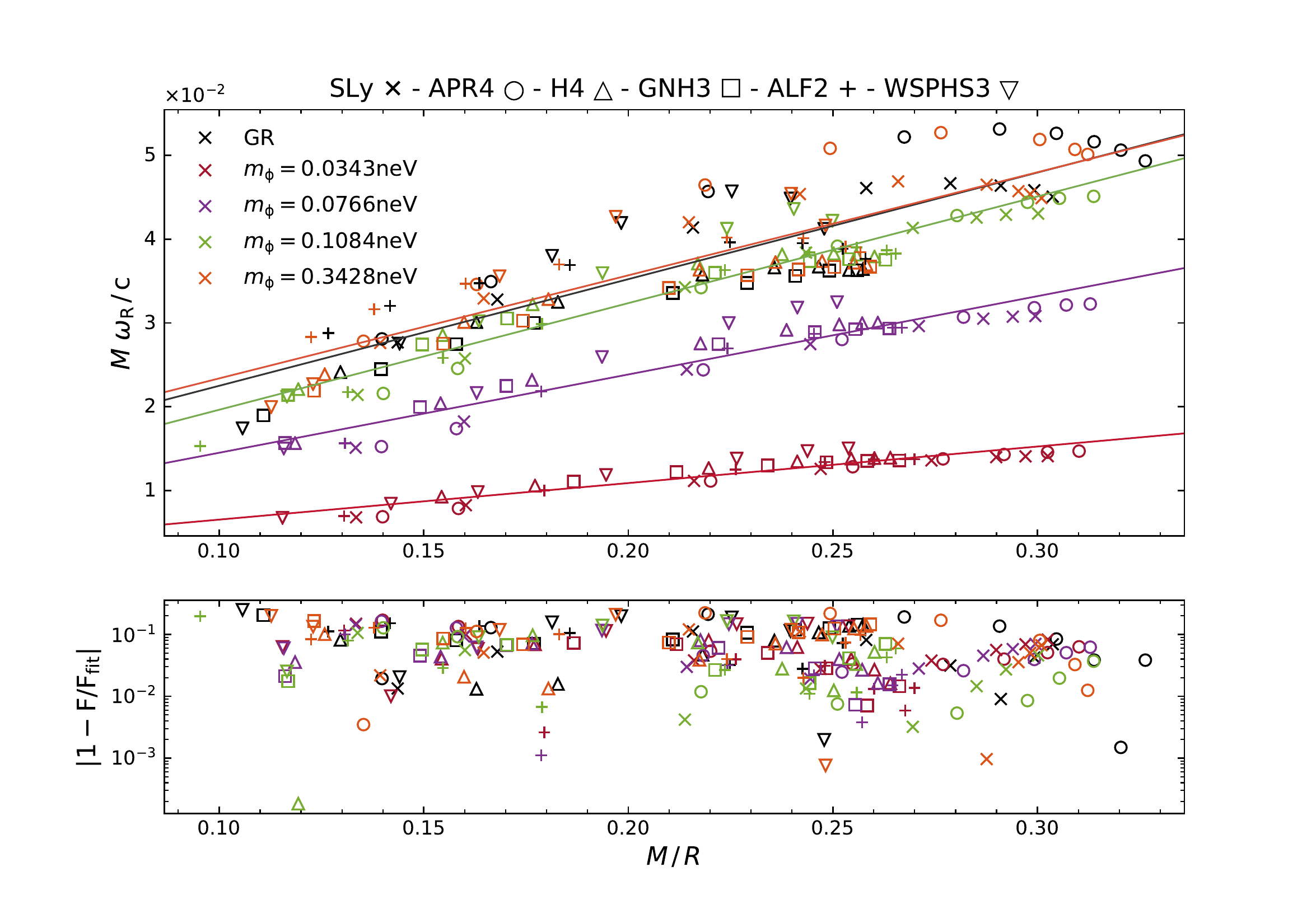}
	\includegraphics[width=.45\textwidth, angle =0]{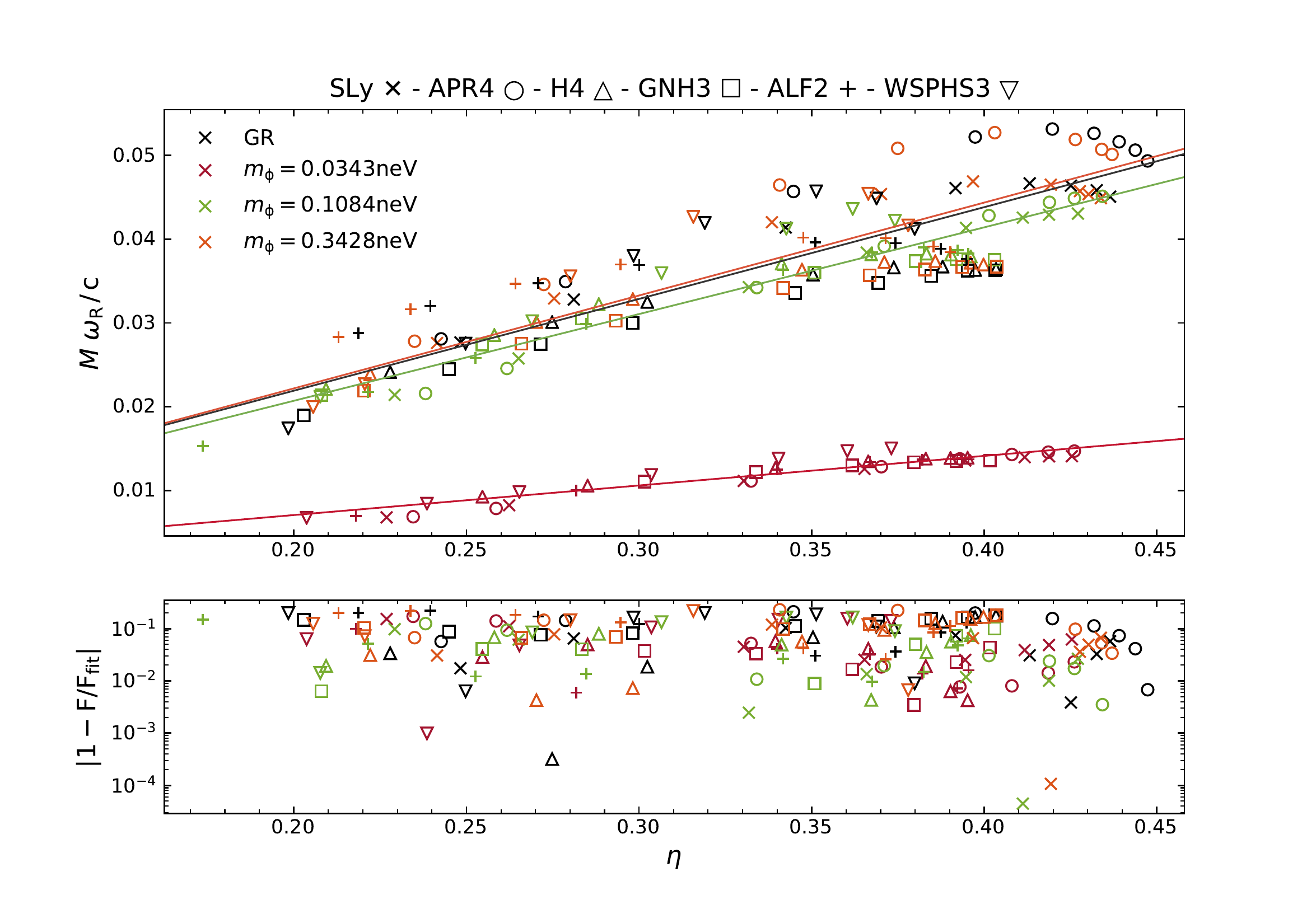}
	\caption{{F-mode universal relations: dimensionless frequency $M\omega_R/c$ (upper panels) and fit errors (lower panels) versus compactness $C=M/R$ (left panels); versus generalized compactness $\eta$ (right panels).
	The symbols indicate the respective EOS and the colors the values of the scalar field mass $m_{\phi}$ with the general relativistic limit in black.}
	}
	\label{fig:uni_rel_MOmegaR_error_eta_linear_affine}
\end{figure}

We now look for universal relations for the fluid dipole F-modes, restricting to relations that involve only the frequency $\omega_R$.
We begin again with the simple scaling relations involving the mass $M$, and exhibit in Figure \ref{fig:uni_rel_MOmegaR_error_eta_linear_affine} the dimensionless frequency $M\omega_R/c$ versus compactness $C=M/R$ (left) and versus generalized compactness $\eta$ (right) together with the fits and error analysis.
The parameters are given in Table \ref{rel_fit_MOmegaR_lhs} for the relation with respect to compactness (left) and for the relation with respect to generalized compactness (right) in Table \ref{rel_fit_MOmegaR_rhs} in Appendix \ref{App_dipole}.

In both cases the relations are not convincing. 
While the errors seem to be largest for general relativity and decrease with decreasing scalar mass $m_\phi$, also the values themselves decrease, and therefore the relative error remains large also for the smallest considered scalar mass, as seen in the lower panels of the figures.

\begin{figure}[h!]
	\centering
\includegraphics[width=.48\textwidth, angle =0]{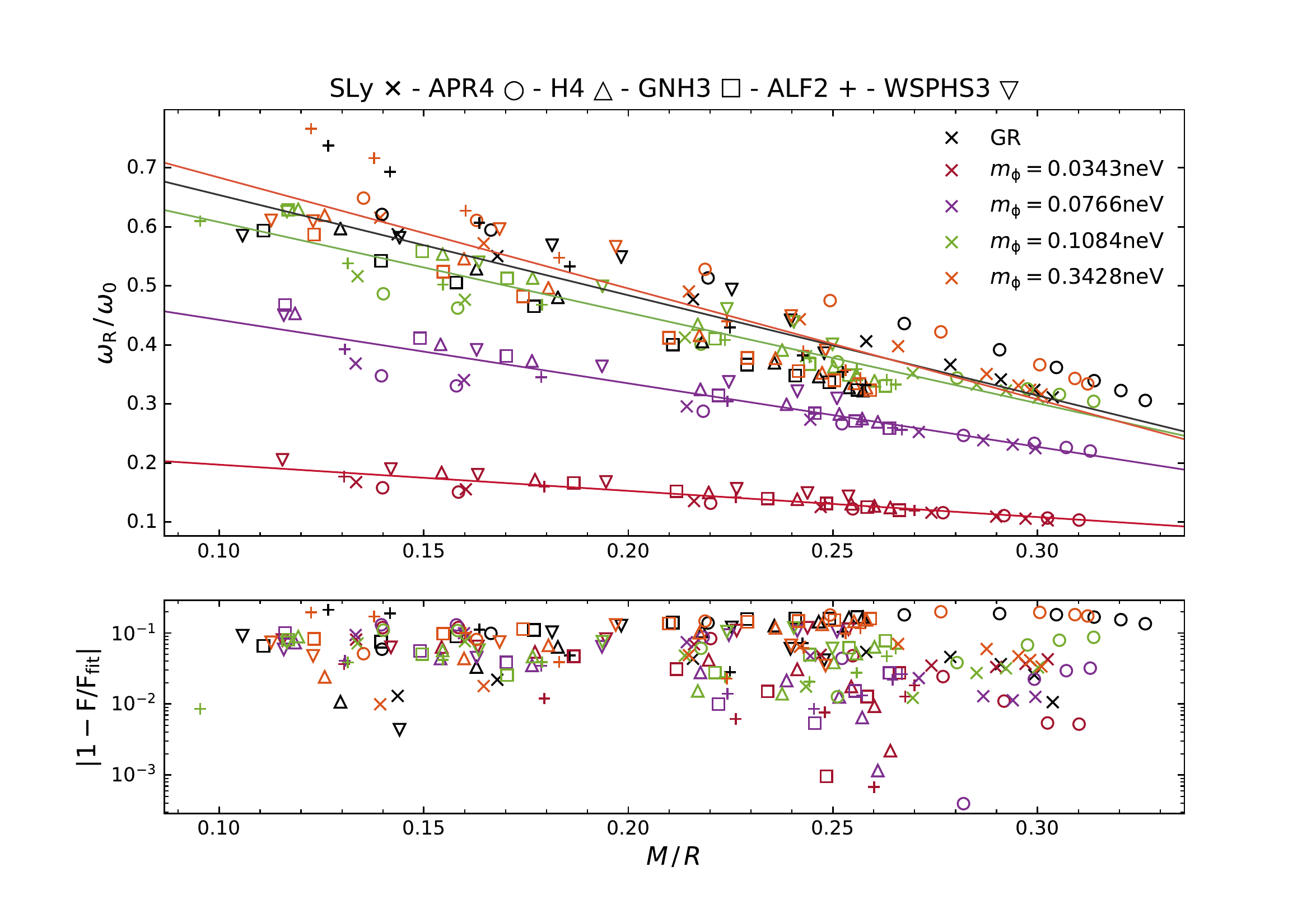}
		\includegraphics[width=.48\textwidth, angle =0]{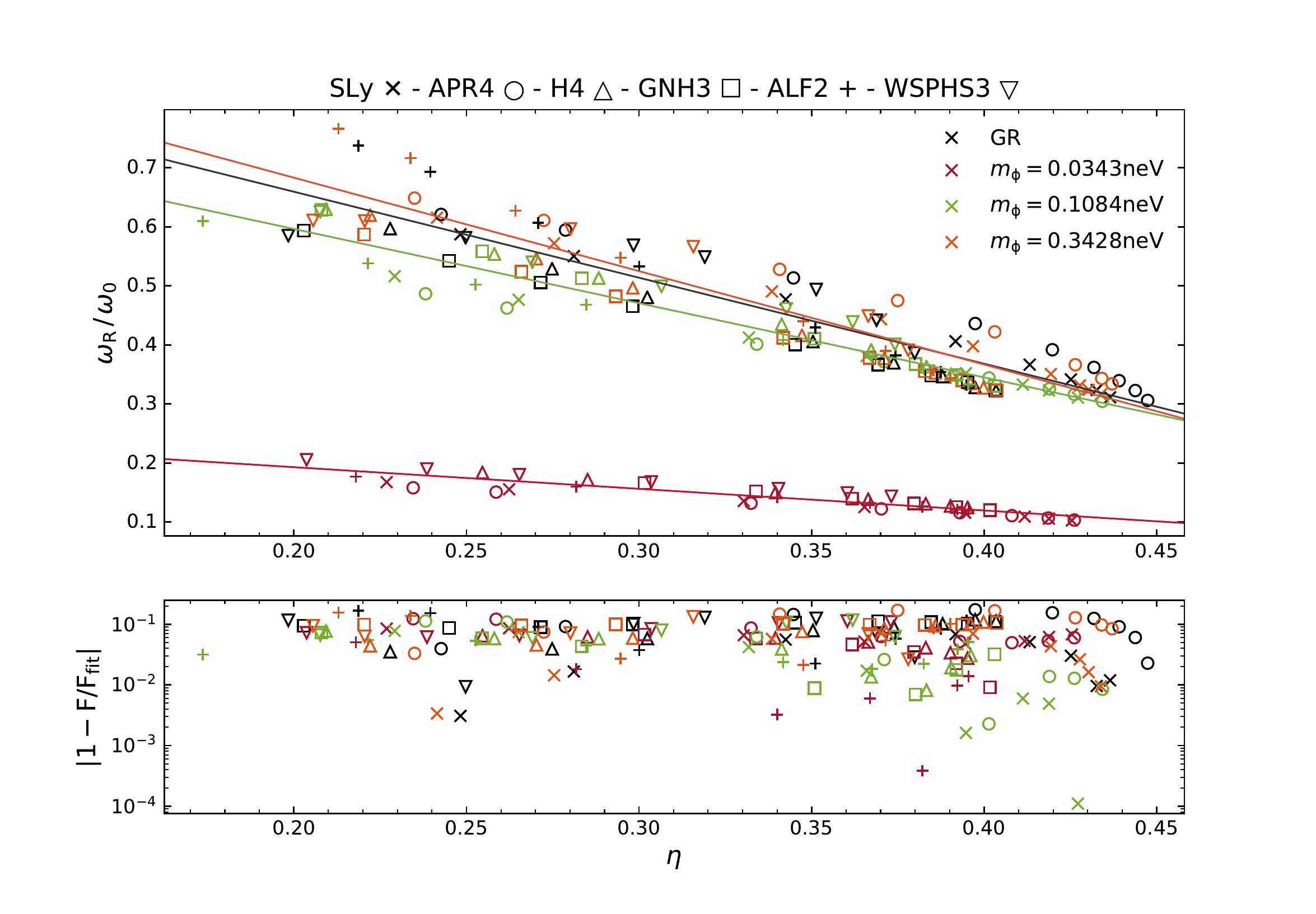}
	\caption{{F-mode universal relations: dimensionless frequency {$\omega_R/\omega_o$} (upper panels) and fit errors (lower panels) versus compactness $C=M/R$ (left panels); versus generalized compactness $\eta$ (right panels).
	The symbols indicate the respective EOS and the colors the values of the scalar field mass $m_{\phi}$ with the general relativistic limit in black.}
		 }
	\label{fig:uni_rel_c_2_l1_error_no_log_linear}
\end{figure}

{Figure \ref{fig:uni_rel_c_2_l1_error_no_log_linear} illustrates the
dimensionless frequency {$\omega_R/\omega_o$} versus the compactness $C=M/R$ (left) and the generalized compactness $\eta$ (right).}
The corresponding fit parameters are given in Table \ref{rel_fit_Omegao_MR} and Table \ref{rel_fit_Omegao_rhs}. 
{In contrast to the f-modes (see Figure \ref{fig:uni_rel_omega_omegao}), the reference frequency $\omega_0$ {in the scaling} provides a small improvement for the F-modes compared to {using} $\hat{\omega}_0$.}
{While these fits for the F-mode are by far not as good as for the f-mode, they are so far the best fits we have obtained for the F-mode.}

Clearly, when looking for universal relations for the F-mode a significant dependence on the EOS is retained, and the considered universal relations are not as good as for other modes.
We also do not see a strong improvement for theories with light scalars as compared to general relativity.
{However, as noted in the previous subsection, for sufficiently small scalar masses universal relations for the F-mode may not be needed, since the frequency $\omega_R$ itself is basically independent of the neutron star mass and, in particular, also independent of the EOS.
{This is further supported by the much lower average error of the fit $\bar{\epsilon}$ {in Table \ref{rel_fit_analysis_l1_lhs}}.}
For larger scalar masses this independence of the EOS (and the neutron star mass) disappears largely, and universal relations would be appreciated just as in the case of general relativity itself, although in the latter case, the modes are non-propagating normal modes.}

\section{Conclusions}

Universal relations of neutron stars are very important as long as their equation of state is not known.
The (approximate) EOS independence of universal relations makes them valuable tools not only for the analysis of the properties of neutron stars in general relativity, but also to put bounds on  generalized gravity theories, when their universal relations deviate from those of general relativity.
In this paper we have for the first time constructed universal relations for polar quasinormal modes in a generalized theory of gravity, beyond the Cowling approximation.
In particular, we have chosen $R^2$ gravity and formulated it in the Einstein frame.
This makes the additional degree of freedom explicit in the form of a scalar field coupled to the neutron star matter, where for the scalar field we have chosen masses in the physically allowed range. 

We have focused our investigations on the fundamental quadrupole f-mode and the dipole F-mode and considered scalar masses in the range $m_\phi=0.0108$ neV to $m_\phi=0.3428$ neV.
The lower of these masses lead to distinct effects as compared to  general relativity, whereas the highest masses yield for these fluid modes results very close to general relativity.
Of course, in general relativity the dipole fluid modes do not propagate, but for the larger scalar masses the F-modes of $R^2$ theory become ultra-long lived, thus approaching the general relativistic limit also with respect to the imaginary part of the mode.

To obtain universal relations we have considered six EOSs, covering plain nuclear matter, nucleons and hyperons, and hybrid nuclear-quark matter.
We have constructed a variety of dimensionless quantities involving the frequency of the modes $\omega_R$ and the damping time $\tau$,
starting with a scaling with the neutron star mass and considering the relations with respect to compactness and generalized compactness.
In the latter the radius has been replaced by the gyration radius, that involves the moment of inertia of the neutron star.
The best universal relations found for the quadrupole f-mode are obtained from scaling with a reference frequency $\hat \omega_o$, that involves the {{{{generalized}}}} compactness $\eta$, versus the compactness $C$, and leads to fit errors smaller than 1\% for the scaled frequency in these $R^2$ theories as well as a significant splitting between these theories.
Future observations of these modes might therefore lead to new constraints on the coupling parameter $a$.

The dipole F-mode of these theories shows a very different behavior.
For the set of scalar masses considered, the frequency $\omega_R$ of the mode itself is very independent of the particular EOS and of the neutron star model employed.
Instead, for a given scalar mass, $\omega_R$ is basically a constant that is determined by the relation $2\pi \omega_R=c/L_\phi$ with Compton wavelength $L_\phi$.
Only for the largest masses considered, deviations occur, and the frequencies of general relativity are approached.
Therefore, for the smaller masses, the frequency itself $\omega_R$ is universal, and a detection would directly yield the scalar mass.
This is very similar to the case of scalar radiation, where we also found this dependency of the frequency on the Compton wavelength \cite{Blazquez-Salcedo:2020ibb,Blazquez-Salcedo:2021exm}.
However, there we did not yet show independence of the EOS, as well.

Our next tasks will involve the calculation and analysis of further quasinormal modes for a representative set of EOSs.
For quadrupole radiation we would like to analyze also the fundamental p mode, the spacetime w-mode and the $\phi$-mode.
For dipole radiation we would like to understand better the $\phi$-mode in $R^2$ theory.
Another direction will be to not only take the moment of inertia into account for the universal relations, but also the tidal deformability and the quadrupole moment of the neutron stars.
Finally, the rapid rotation case will have to be tackled for quasinormal modes in alternative gravities.

\section{Acknowledgements}

{We would like to gratefully acknowledge support by the DFG Research Training Group 1620 \textit{Models of Gravity}, DFG projects BL1553 and Ku612/18-1, FCT project PTDC/FIS-AST/3041/2020, and the COST Actions CA15117 and CA16104. FSK thanks the Department of Theoretical Physics and IPARCOS of the Complutense University of Madrid for their hospitality.
}


\appendix
\section{Perturbation equations for the dipole ($l=1$) modes.}\label{Appendix_eqs_l1}

Here we present the minimal set of perturbation equations for the dipole mode. The quantities
$\hat{Q}$, $\hat{\rho}$ and $\hat{p}$ are defined in the Jordan frame,
with $\hat{Q} = \frac{d\hat{\rho}}{d\hat{p}}$.
{For convenience we define}
\begin{eqnarray}
A_1 &=& 1+{{ e}^{-{\frac {4}{\sqrt {3}}\Phi_0 }}} -2{{e}^{-\frac{2}{\sqrt {3}}\Phi_0  }} \ ,
 \nonumber \\
A_2 &=& {{ e}^{-{\frac {4}{\sqrt{3}}\Phi_0}}}-{{ e}^{-{\frac {2}{\sqrt{3}}\Phi_0}}} \ .
\end{eqnarray}

{Following \cite{Campolattaro:1970}, we fix the gauge by taking}
\begin{eqnarray}
K = 0 \ .
\end{eqnarray}

{We obtain the following set of differential equations for the perturbation functions.
From the scalar field equation:}
\begin{align} 
 &   {\frac {{d}^{2}\Phi_1}{{d}{r}^{2}}}  = \left( 
 \left( {\frac {{{e}^{2\lambda}} \sqrt {3}}{16}
   } \left( -{\frac { \hat{Q}  }{3}   }
+ 1 \right) {H_2} A_1 \right. 
 \left.
 +
 \left( {\frac { \hat{Q}  }{3}  }- 
 {1  } \right) \frac{1}{4} {
{ e}^{2\lambda  }}  \Phi_1  A_2 
 \right)    \right.
 \nn 
 \\& \left.
 -
 {\frac {  {{ e}^{2\lambda  }}  r  }{16} A_1  }   
 \left(
 H_0 - \hat{Q} H_2
 \right) \left( {\frac { d \Phi_0 }{{ d}r}} 
 \right)
 + \left( -{\frac {r\sqrt {3}
  { { e}^{2\lambda  }} \hat{Q}
  }{12}  {\frac { d\Phi_0 }{{ d}r}}   A_2  }   \right.  \right.
\nn 
\\ &   \left.  \left.
+ \frac {  {{ e}^{2\lambda }} \sqrt {3}}{36} \left( {{ e}^{-\frac{4}{\sqrt {3}} \Phi_0
  }} \left(-3\, \left( {\frac 
{ d \Phi_0 }{{ d}r}}  \right) r
+ 4\,\sqrt {3} \right)
\right.  \right.  
 \left.  \left.
+ {{ e}^{-\frac{2}{\sqrt {3}}\Phi_0
  }} \left(  3\, \left( {\frac { d\Phi_0 }{{ d}r}
}  \right) r
 -2\,\sqrt {3} \right) \right)  \right) \Phi_1  \right) \alpha
 \nn 
 \\ &
 + \frac {16\,\pi\, {{ e}^{2\lambda
  }} \hat{\rho} }{3} \left( \hat{\rho}-\hat{p} \hat{Q} \right) \Phi_1  {
{ e}^{-{\frac {4 }{\sqrt {3}}}\,\Phi_0  }}
\nn 
\\ &
+ 
{\frac {\sqrt {3}}{2\,  {{ e}^{2\nu }}   }}
\left(  \left( {\frac {\hat{Q}
  }{3}}-1 \right) {H_1}  {
\omega}^{2}  
+ 
\left(  \left( {\frac { d\nu}{{ d}r}}
   \right) r+  {{ e}^{2\lambda }} -1 \right)
{\frac { \sqrt {3} }{2\,{r}^{2}}}
\left( -{\frac { \hat{Q}  }{3
}}
+
1 \right) {H_0}  \right.
\nn 
\\ & \left.
+ \left( { \left(\frac {\sqrt {3}\hat{Q} }{6\,{r}^{2}}
-
\frac {\sqrt {3}}{2\,{r}^{2}}
\right)
\left( 8\,\pi\,{{ e}^{-\frac{4}{\sqrt {3}} \,
\Phi_0  }} {{ e}^{2\lambda  }} \hat{p} {r}^{2}- \left( {\frac { d\nu}{
{ d}r}}   \right) r+  {{ e}^{2\lambda
  }}  -1 \right) }  \right) {H_2}  \right.
 \nn 
 \\ &  \left.
 + \left(  
16\,r^2 \pi\,{{ e}^{-\frac{4}{\sqrt {3}} \,\Phi_0  }}
 {{ e}^{2\lambda  }}  (\hat{p}\hat{Q} - \hat{\rho})
  +3 \hat{Q}  - 9  
   \right) \frac { \Phi_1 }{ \sqrt {3} \,r}  {\frac { d\Phi_0}{{ d}r}}  \right. 
 \nn 
 \\ &  \left.
 + \sqrt {3}
 \left( {\frac { d\Phi_1}{{ d}r}}  \right) {\frac { d\Phi_0 }{{ d}r}} \left(
 {\frac {  \hat{Q}
  }{3}}- 1 \right) \right) 
 + \left(
  {\frac { d \Phi_0  }{{ d}r}}
 \frac { r}{2\, {{ e}^{2\nu  }}  }
 \left( -\hat{Q}+1
 \right) {H_1}  -{\frac { {{ e}^{
2\lambda  }}  \Phi_1  }{
  {{ e}^{2\nu}} }} \right) {
\omega}^{2}
\nn 
\\ &
+ \left(   {\frac { d \Phi_0}{{ d}r}}
    \left(  \left( {\frac { d\nu}{{ d}r}}
   \right) r+ {{ e}^{2\lambda }} -1 \right) {\frac {\hat{Q}  }{2r} } 
 -
{\frac { d\Phi_0}{{ d}r}}  \frac{1}{2\,r} \left( 
8\,{r}^{2}\pi\,{{ e}^{-\frac{4}{\sqrt {3}} \,\Phi_0  
}} {{ e}^{2\lambda  }} \hat{\rho}
  \right . \right.
 \nn 
 \\ & \left .\left.
  + \left( {\frac { d\Phi_0}{{ d}r}}  \right) ^{2}{r}^{2}
  + \left( {\frac { d\nu}{{ d}r}}
   \right) r 
 -2\, \left( {\frac { d\lambda }{{ d}r}}
  \right) r-  {{ e}^{2\lambda }} \vphantom{} \right) \right) { H_0}  
 \nn 
 \\ &
 + \left( -{\frac {\hat{Q}   }{2\,r} {\frac { d \Phi_0}{
{ d}r}} \left( 8\,\pi\,{{ e}^{-\frac{4}{\sqrt {3}}\,\Phi_0  }}  {{ e}^{2\lambda
  }} \hat{p} {r}^{2}-
 \left( {\frac { d\nu}{{ d}r}}   \right) r+
 {{ e}^{2\lambda  }}  -1 \right) }  \right.
 \nn 
 \\ &  \left. 
-{
\frac{1}{2r} \left(
 \left( {\frac { d\Phi_0 }{{ d}r}} 
 \right) ^{3}{r}^{2}+ \left( {\frac { d\Phi_0 }{{ d}r}} \right) {{ e}^{2\lambda}} - \left( {\frac { d\nu}{{ d}r}}  
 \right)  \left( {\frac { d\Phi_0 }{{ d}r}}  
 \right) r-2\,r{\frac {{ d}^{2}\Phi_0 }{{ d}{r}^{2}}} -2\,{\frac { d\Phi_0 }{{ d}r}}   \right)}
 \right) { H_2}  
 \nn 
 \\ &
 + \left( -3\, \left( {\frac 
{ d\Phi_0 }{{ d}r}}   \right) ^{2}\hat{Q} +
\frac{1}{r^2} \left(
{3\, \left( {\frac { d\Phi_0}{{ d}r}}  \right) ^{2}{r}^{2}+ \left( {\frac { d\lambda}{{ d}r}}
   \right) r- \left( {\frac { d\nu}{{ d}r}}
   \right) r+2\, {{ e}^{2\lambda }} -2 } \right) \right) \Phi_1  
 \nn 
 \\ &
 -
r \left( {\frac { d\Phi_1}{{ d}r}}   \right) \hat{Q}   \left( {\frac { d\Phi_0}{{ d}r}}  \right) ^{2}+
\frac{1}{r}
\left(  \left( {\frac { d\Phi_0  }{{ d}r
}} \right) ^{2}{r}^{2}+ \left( {\frac { d\lambda}{
{ d}r}}   \right) r- \left( {\frac { d\nu}{
{ d}r}}   \right) r-4 \right) {\frac { d\Phi_1 }{
{ d}r}} \, .
\end{align}

{From the Einstein equations we get the following set of equations:}
\begin{align}
    {\frac { d H_0}{{ d}r}} =  -{\frac {r{
\omega}^{2}{H_1}  }{  {{ e}^{2\nu}}  }}- \left( {\frac { d \nu}{{ d}r}}
   \right) {H_0}  
-
\frac{1}{r}
{ \left(  \left( {\frac { d \nu  }{{ d}r}}
 \right) r+1 \right) {H_2}   }
 +4\,\Phi_1
  {\frac { d \Phi_0  }{{ d}r}} \, ,
\end{align}

\begin{align}
  &  {\frac { d {H_1}}{{ d}r}}  = -\frac {
  {{ e}^{2\lambda  }}  {
{ e}^{2\nu  }}  {H_2} \alpha}{4\,r{\omega}^{2}}
A_1
 + \left( -{\frac { d\nu }{{ d}r}}
 +{\frac { d\lambda }{{ d}r}} \right) {H_1}  +{\frac {{
H_2} {{ e}^{2\lambda}} }{r}}
\nn 
 \\ &
+\frac { 
{ e}^{2\nu  } }{r{\omega}^{2}} 
\left[ {H_0}  
\left(  \left( {\frac 
{ d\nu}{{ d}r}}   \right) ^{2} - \left( {
\frac { d\nu}{{ d}r}}   \right)  \left( {\frac 
{ d\lambda}{{ d}r}}   \right) + \left( {
\frac {{ d}^{2} \nu}{{ d}{r}^{2}}}   \right) - \frac{1}{r} \left( {\frac { d\nu}{ { d}r}}   \right)  \right. \right.
\nn
\\ &  \left. \left.
+
 \frac{1}{r}\left( {\frac { d \lambda}{{ d}r}}   \right) -
 \frac{{{ e}^{2\lambda  }} }{r^2}+ \frac{1}{r^2} \right) 
+ {H_2} \left( 16\,\pi\,{{ e}^{-\frac{4}{\sqrt {3}}\,\Phi_0 \left( 
r \right) }} {{ e}^{2\lambda  }}
\hat{p}   \right. \right.
\nn 
\\ & \left. \left.
- \left( {\frac { d \nu}{{ d}r}}
   \right) ^{2}+ \left( {\frac { d \nu}{{ d}r
}}   \right)  \left( {\frac { d\lambda  }{{ d}r}}
 \right) -2\, \left( {\frac { d\Phi_0 }{
{ d}r}}  \right) ^{2}- \left( {\frac {
{ d}^{2}\nu}{{ d}{r}^{2}}}   \right)   \right. \right.
\nn 
\\ & \left. \left.
+
\frac{1}{r^2} \left(
 {{ e}^{2\lambda  }} - \left( {
\frac { d\nu  }{{ d}r}} \right) r+ \left( {
\frac { d\lambda }{{ d}r}}  \right) r-1 \right)\right)
+{4}\, \left( {\frac { d\Phi_0}{{ d}r}}   \right) {\frac { d\Phi_1 }{{ d}r}}
\right.
\nn 
\\ &  \left.
+ 4 \,
 \left( - \left( {\frac { d\nu }{{ d}r}} 
 \right)  \left( {\frac { d\Phi_0  }{{ d}r}}
 \right) + \left( {\frac { d\Phi_0  }{{ d}r}}
 \right)  \left( {\frac { d\lambda  }{{ d}r}}
 \right) - {\frac {{ d}^{2}\Phi_0}{{ d}{r}^{2}}} + \frac{1}{r} {\frac { d\Phi_0}{{ d}r}}   \right) 
\Phi_1  \right]  \, ,
\end{align}

\begin{align}
 &   {\frac { d H_2}{{ d}r}}  = \alpha r {{ e}^{2\lambda  }}  \left( {
\frac { A_1 }{8} \left( -H_0 + \hat{Q} H_2 \right) } 
 -  {\frac {\sqrt {3} \Phi_1 A_2  }{6}
 \left( \hat{Q} + 1 \right) 
} 
  \right) 
\nn 
 \\ &
+ \left( {\frac {32 \,r\pi\, {{ e}^{2\lambda }}  }{\sqrt {3}}{
{ e}^{-{\frac {4 }{ \sqrt {3} }} \,\Phi_0}}}
(-\hat{\rho}+ \hat{p} \hat{Q} )
 \right) \Phi_1  
 -{\frac {r\hat{Q} {\omega}^{2}{ H_1}  }{ {
{ e}^{2\nu  }}  }}
\nn 
 \\ &
+ \frac{{H_0}  }{r} \left(  \hat{Q}
 \left(  \left( {\frac { d\nu}{{ d}r}}  
 \right) r+ {{ e}^{2\lambda  }} -
1 \right)  \right.
\nn 
 \\ &  \left. 
{- 8\,{r}^{2}
\pi\,{{ e}^{-\frac{4}{\sqrt {3}}\,\Phi_0  }} {{ e}^{2\lambda  }}\hat{\rho} - \left( {\frac { d\Phi_0  }{{ d}r}}
 \right) ^{2}{r}^{2}+2\, \left( {\frac { d\lambda}{{ d}r}}
   \right) r+ {{ e}^{2\lambda }} -1  } \right) 
 \nn 
 \\ &
 + {H_2} \left( -{\frac {\hat{Q}  }{r} \left( 8\,\pi\,
{{ e}^{-\frac{4}{\sqrt {3}}\,\Phi_0  }} {
{ e}^{2\lambda}}\hat{p}{r}^{2}- \left( {\frac { d\nu}{{ d}r}}  \right) r+ {{ e}^{2\lambda  }} -1 \right) }  \right.
 \nn 
 \\ &  \left. 
 -  \left( {\frac { d\Phi_0 }{{ d}r}}
  \right) ^{2}{r}+2\, \left( {\frac { d\lambda }{
{ d}r}}  \right)  -
\frac{ {{ e}^{
2\lambda  }} +2}{r} \right) 
  + 2\,{\frac { d\Phi_0}{{ d}r}} \left( 1- 3\hat{Q}  
   \right) \Phi_1 
 \nn 
 \\ &
 -2r\, \left( {\frac 
{ d\Phi_0 }{{ d}r}}  \right)  \left( {\frac 
{ d\Phi_1  }{{ d}r}} \right) (\hat{Q} - 1) \, , 
\end{align}

\begin{align}
    {\Pi}_1  = & \left( {\frac {{H_2}
  }{64\,\pi}  A_1  }  
 -\frac {\sqrt {3}\Phi_1  }{
48\,\pi}  A_2  \right) \alpha-{\frac {{\omega}^{2}{H_1} }{8\,  {{ e}^{2\lambda  }} {{ e}^{2\nu  }}  \pi}}
+
\left(  \left( {\frac { d\nu}{{ d}r}} 
 \right) r+  {{ e}^{2\lambda  }} -
1 \right)
{\frac { { H_0}  }{8\,\pi\,{r}^{2}  {
{ e}^{2\lambda  }}  }} 
\nn 
 \\ &
-{\frac {{H_2}  }{8\,\pi\,{r}^{2} {{ e}^{2\lambda
  }}  } \left( 8\,\pi\,{{ e}^{-\frac{4}{\sqrt {3}}\,\Phi_0  }} {{ e}^{2\lambda
  }}\hat{p} {r}^{2}-
 \left( {\frac { d\nu}{{ d}r}}   \right) r+
 {{ e}^{2\lambda  }} -1 \right) }
 \nn 
 \\ &
-\frac{1} {{4\, {{ e}^{2\lambda  }} \pi}} \left( {\frac { d\Phi_0  }{{ d}r}}
 \right) \left({\frac { d\Phi_1}{{ d}r}}   + {\frac {3\,
\Phi_1   }{r}} \right)   \, ,
\end{align}

\begin{align}
    {\it E}_1  = & \left( {\frac {  {H_2}  }{64\,\pi}  A_1  } 
\right.
\left. 
-{\frac {\sqrt {3}
  \Phi_1  }{48\,\pi}  A_2  } \right) \alpha \hat{Q}
 +  {\frac {4}{\sqrt {3} }{{ e}^{-{\frac {4}{\sqrt {3}}}\,\Phi_0 }}}
 \left( \hat{Q} \hat{p} - \hat{\rho} 
 \right) \Phi_1  
 \nn 
\\ &
 -{\frac {\hat{Q}  {
\omega}^{2}{H_1}  }{8\,  {{ e}^{
2\lambda  }}  {{ e}^{2\nu }} \pi}}
+
\left(  \left( {\frac { d\nu}{
{ d}r}}   \right) r+ {{ e}^{2\lambda
  }}  -1 \right) 
{\frac { \hat{Q}  {H_0}  }{8\,\pi\,{r}^{2} {{ e}^{2\lambda  }}  }}
-{\frac {  \hat{Q}   }{4\, {{ e}^{2\lambda
  }} \pi}}
\left(  {\frac { d\Phi_1}{
{ d}r}} + \frac{3\Phi_1}{r}
\right) {\frac { d\Phi_0 }{{ d}r}}
\nn 
\\ &
-{\frac {\hat{Q} {H_2}  }{8\,\pi\,{r}^{2} {
{ e}^{2\lambda  }} } \left( 8\,\pi\,{
{ e}^{-\frac{4}{\sqrt {3}}\,\Phi_0  }} {
{e}^{2\lambda  }}\hat{p}{r}^{2}- \left( {\frac { d\nu}{{ d}r}} \right) r+  {{e}^{2\lambda}} -1 \right) } \, ,
\end{align}

\begin{align}
 &    {\it V}  =  \left({\frac { {{ e}^{2\nu
  }} {H_2}  
\alpha}{64\,\pi\, \left( \hat{p} +\hat{\rho}\right) {\omega}^{2}}  A_1 }
 -{\frac {{H_1}}{8\,\pi\, \left( \hat{p} +\hat{\rho} \right) {{ e}^{2\lambda }} } }
 \right) {{ e}^{{\frac {4 }{\sqrt {3}}}\,
\Phi_0 }}
 \nn 
 \\ &
 +\frac {1}{{\omega}^{2}} \left[ -\frac { {
{ e}^{2\nu  }}
}
{16\,\pi\,{r}
^{2} \left( \hat{p} +\hat{\rho} 
 \right)  {{ e}^{2\lambda  }} }
\left( - \left( {\frac 
{ d\lambda}{{ d}r}}   \right)  \left( {\frac 
{ d\nu}{{ d}r}}   \right) {r}^{2}+ \left( {
\frac { d\nu}{{ d}r}}   \right) ^{2}{r}^{2} \right. \right.
\nn 
\\ &   \left.\left.
+
 \left( {\frac {{ d}^{2}\nu}{{ d}{r}^{2}}} 
 \right) {r}^{2}+ \left( {\frac { d\lambda }{{ d}r}} \right) r- \left( {\frac { d\nu}{{ d}r}} \right) r- {{ e}^{2\lambda  }} +1 \right) {H_0}   
 \right.
\nn 
\\ & \left.
-
\frac { {{ e}^{2\nu  }} {H_2}  }{16\,\pi\,{r}^{2} \left( \hat{p}+\hat{\rho}  \right)  {{ e}^{2\lambda}} } \left( 16\, {{ e}^{
2\lambda  }}  \hat{p} \pi\,{r
}^{2} {{ e}^{-{\frac {4 }{\sqrt {3}}}
\,\Phi_0 }}
+\left( {\frac { d\lambda}{{ d}r}}  \right) 
 \left( {\frac { d\nu}{{ d}r}}   \right) {r}^{2}   \right. \right.
 \nn 
 \\ &  \left. \left.
-
 \left( {\frac { d\nu}{{ d}r}}   \right) ^{2}{r}
^{2}-2\,
 \left( {\frac { d\Phi_0}{{ d}r}}   \right) ^{2}{
r}^{2}- \left( {\frac {{ d}^{2}\nu}{{ d}{r}^{2}}}  
 \right) {r}^{2} \right. 
 \left.
 +  {{ e}^{2\lambda  }}
+ \left( {\frac { d\lambda}{{ d}r}}   \right) r
-
 \left( {\frac { d\nu}{{ d}r}}   \right) r- 1
 \right) \right. 
 \nn 
 \\ &  \left. 
 -{\frac { {{ e}^{2\nu  }}  \Phi_1  }{4\,
\pi\,r \left( \hat{p} +\hat{\rho} 
 \right)  {{ e}^{2\lambda  }}  }}  
 \left( - \left( {\frac { d\Phi_0  }{{ d}r}}
 \right) r{\frac { d\nu}{{ d}r}}  + \left( {
\frac { d\Phi_0 }{{ d}r}}  \right)  \left( {
\frac { d\lambda}{{ d}r}}   \right) r-r{\frac {
{ d}^{2}\Phi_0}{{ d}{r}^{2}}}  +{\frac { d\Phi_0 }{
{ d}r}}  \right) 
 \right.
\nn 
 \\ & \left. 
-
\frac { {{ e}^{2\nu  }} }{4\,\pi\, \left( \hat{p} +\hat{\rho}  \right)  {{ e}^{2\lambda}} } 
 \left( {
\frac { d\Phi_0 }{{ d}r}}  \right) {\frac { d \Phi_1}
{{ d}r}} 
\right]  
{{ e}^{{\frac {4}{\sqrt {3}}}
\,\Phi_0  }} \, ,
\end{align}

\begin{align}
     {\it W}  = & 
    \frac{ {{ e}^{{\frac {4  }{\sqrt {3}}}\,\Phi_0 }} }{4\,\pi\,
 \left( \hat{p} +\hat{\rho}  \right) {
{ e}^{\lambda  }}}
     \left(
     {\frac {{r}^{2}\alpha\,{
H_1}  }{16}
 A_1 } +
 {\frac {{H_2} }{2} } 
 -
{\Phi_1   \left( {
\frac { d\Phi_0}{{ d}r}}   \right) r}  \right.
\nn
\\ & \left. 
+\frac {  {H_1}  }{2 {{ e}^{2\lambda  }}  } \left( 8\,\hat{\rho}  {r}^{2}\pi\, {{ e}^{2\lambda  }}
 {{ e}^{{
-\frac {4  }{\sqrt {3}}\,\Phi_0 }}}
 + \left( {\frac { d\Phi_0}{{ d}r}}  \right) ^{2}{r}^{2} \right. 
\left.
 -2 \left( {\frac { d\lambda }{{ d}r}} \right) r+
 1 \right) 
 \right) \, .
\end{align}

%

\section{Tables for universal relations for $l=2$ f-mode.}\label{Appendix_tables_l2}

{This section presents the tables of the fit parameters and their errors for the figures of the quadrupole f-mode universal relations in section \ref{Uni_rel_l2}.} The last column of each table represents the average error $\bar{\epsilon}$, which is calculated for each theory as follows
\begin{equation}
    \bar{\epsilon} = \frac{1}{N}\sum\limits_{k=1}^N \left|1-\frac{F_k}{F_{\mathrm{fit,}k}}\right|,
\end{equation}
where $N$ is the total number of points for each theory.
\begin{table}[H]
  \centering
  \begin{tabular}{|c|c|c|c|c|c|c|}
    \hline
    \text{Theory} & $a$ & $b$ & $c$ & $d$ & $e$ & $\bar{\epsilon} [\%]$\\
    \hline
    \hline 
    GR &   $ -1 \pm 3$ 
    & $ -0.1 \pm 3 $ 
    &   $ 0.4 \pm 0.9$ 
    &    $ -0.005 \pm 0.1$ 
    &  $ 0.003 \pm 0.006 $  
    & $1.5$\\
    \hline 
      $m_{\phi}=0.343$neV &  $-2 \pm 5$ 
    & $ 0.2 \pm 5$ 
    &   $0.4 \pm 1$ 
    &  $-0.0006 \pm 0.2$ 
    & $0.003 \pm 0.009 $ 
    & $1.7$\\
     \hline 
    $m_{\phi}=0.108$neV &  $  -7 \pm 6 $ 
    & $ 5 \pm 5 $ 
    &   $ -1 \pm 2 $ 
    &  $ 2 \pm 0.2 $ 
    & $  -0.005 \pm 0.01  $ 
    & 1.6\\
     \hline 
      $m_{\phi}=0.0343$neV &  $-7 \pm 5 $ 
    & $ 5 \pm 5 $ 
    &   $-1 \pm 1 $ 
    &  $ 0.2 \pm 0.2 $ 
    & $ -6 \pm 0.009 $ 
    & 1.6\\
     \hline 
      $m_{\phi}=0.0108$neV &  $ -8 \pm 5$ 
    & $5 \pm 5 $ 
    &   $-1 \pm 1 $ 
    &  $ 0.2 \pm 0.2 $ 
    & $ -0.007 \pm 0.009 $
    & 1.5\\
     \hline 
  \end{tabular}
  {\caption{Fit parameters and their respective errors for Figure \ref{fig:uni_rel_2_mass_omega} (left)} \label{rel_fit_MOmegaR_C_l2_lhs}}
\end{table}

\begin{table}[H]
  \centering
  \begin{tabular}{|c|c|c|c|c|c|c|}
    \hline
    \text{Theory} & $a$ & $b$ & $c$ & $d$ & $e$ & $\bar{\epsilon} [\%]$\\
    \hline
    \hline 
    GR &   $ 0.3 \pm 0.3$ 
    & $ -0.4 \pm 0.4 $ 
    &   $ 0.3 \pm 0.2$ 
    &    $ -0.02 \pm 0.04$ 
    &  $ 0.002 \pm 0.003$
    & 0.2\\
    \hline 
      $m_{\phi}=0.343$neV &  $0.4 \pm 0.4$ 
    & $-0.5 \pm 0.5$ 
    &   $0.3 \pm 0.2$ 
    &  $-0.02 \pm 0.05$ 
    & $0.002 \pm 0.004 $
    & 0.2\\
     \hline 
    $m_{\phi}=0.108$neV &  $  0.2 \pm 0.4 $ 
    & $ -0.3 \pm 0.5 $ 
    &   $ 0.2 \pm 0.2 $ 
    &  $ -0.003 \pm 0.04 $ 
    & $  0.001 \pm 0.003  $
    & 0.2\\
     \hline 
      $m_{\phi}=0.0343$neV &  $0.5 \pm 0.3 $ 
    & $-0.7 \pm 0.3 $ 
    &   $0.4 \pm 0.2 $ 
    &  $-0.04 \pm 0.03 $ 
    & $0.004 \pm 0.002 $
    & 0.1\\
     \hline 
      $m_{\phi}=0.0108$neV &  $0.4 \pm 0.3$ 
    & $-0.5 \pm 0.4$ 
    &   $0.3 \pm 0.2 $ 
    &  $-0.03 \pm 0.04 $ 
    & $0.003 \pm 0.003 $
    & 0.2\\
     \hline 
  \end{tabular}
  {\caption{Fit parameters and their respective errors for Figure \ref{fig:uni_rel_2_mass_omega} (right)} \label{rel_fit_MOmegaR_eta_l2_rhs}}
\end{table}

\begin{table}[H]
{\small 
  \centering
  \begin{tabular}{|c|c|c|c|c|c|c|}
    \hline
    \text{Theory} & $a$ & $b$ & $c$ & $d\cdot10^{-3}$ & $e\cdot10^{-4}$ & $\bar{\epsilon} [\%]$\\
    \hline
    \hline 
    GR &   $ -0.01 \pm 0.01$ 
    & $ -0.011 \pm 0.009 $ 
    &   $ 0.009 \pm 0.003 $ 
    &    $ -1.3 \pm 0.4 $ 
    &  $ 0.6 \pm 0.2$ 
    & $2.1$\\
    \hline 
      $m_{\phi}=0.343$neV &  $-0.02 \pm 0.02$ 
    & $-0.006 \pm 0.02$ 
    &   $0.007 \pm 0.005$ 
    &  $-1.1 \pm 0.6$ 
    & $0.5 \pm 0.3 $
    & $2.2$\\
     \hline 
    $m_{\phi}=0.108$neV &  $  -0.03 \pm 0.02 $ 
    & $ 0.0008 \pm 0.02 $ 
    &   $ 0.005 \pm 0.006 $ 
    &  $ -0.8 \pm 0.7 $ 
    & $  0.4 \pm 0.4  $  
    & $2.1$\\
     \hline 
      $m_{\phi}=0.0343$neV &  $ -0.03 \pm 0.02 $ 
    & $ 0.005 \pm 0.02 $ 
    &   $0.004 \pm 0.005 $ 
    &  $-0.7 \pm 0.7 $ 
    & $0.3 \pm 0.3 $
    & $1.8$\\
     \hline 
      $m_{\phi}=0.0108$neV &  $ -0.05 \pm 0.02$ 
    & $ 0.02 \pm 0.02$ 
    &   $0.0009 \pm 0.005 $ 
    &  $-0.2 \pm 0.7 $ 
    & $0.1 \pm 0.3 $
    & $1.8$\\
     \hline 
  \end{tabular}
  {\caption{Fit parameters and their respective errors for Figure \ref{fig:uni_rel_mass_tau_full} (left)} \label{rel_fit_MOmegaI_l2_lhs}}
  }
\end{table}

\begin{table}[H]
{\small  
  \centering
  \begin{tabular}{|c|c|c|c|c|c|c|}
    \hline
    \text{Theory} & $a$ & $b$ & $c$ & $d\cdot10^{-3}$ & $e\cdot10^{-4}$ & $\bar{\epsilon} [\%]$\\
    \hline
    \hline 
    GR &   $ -0.062 \pm 0.003$ 
    & $ 0.067 \pm 0.004 $ 
    &   $ -0.026 \pm 0.002 $ 
    &    $ 4.5 \pm 0.4$ 
    &  $ -2.9 \pm 0.3 $
    & $1.0$\\
    \hline 
      $m_{\phi}=0.343$neV &  $ -0.059 \pm 0.004$ 
    & $ 0.063 \pm 0.005$ 
    &   $ -0.024 \pm 0.003$ 
    &  $ 4.1 \pm 0.5$ 
    & $ -2.7 \pm 0.4 $
    & $1.0$\\
     \hline 
    $m_{\phi}=0.108$neV &  $  -0.058 \pm 0.004 $ 
    & $ 0.060 \pm 0.006 $ 
    &   $ -0.022 \pm 0.003 $ 
    &  $ 3.7 \pm 0.5 $ 
    & $  -2.4 \pm 0.4  $
    & $0.9$\\
     \hline 
      $m_{\phi}=0.0343$neV &  $ -0.059 \pm 0.003 $ 
    & $ 0.060 \pm 0.004 $ 
    &   $ -0.021 \pm 0.002 $ 
    &  $ 3.6 \pm 0.4 $ 
    & $ -2.2 \pm 0.3 $
    & $0.6$\\
     \hline 
      $m_{\phi}=0.0108$neV &  $ -0.063 \pm 0.003$ 
    & $ 0.065 \pm 0.004 $ 
    &   $ -0.023 \pm 0.002 $ 
    &  $ 4.0 \pm 0.4 $ 
    & $-2.5 \pm 0.3 $
    & $0.6$\\
     \hline 
  \end{tabular}
  {\caption{Fit parameters and their respective errors for Figure \ref{fig:uni_rel_mass_tau_full} (right)} \label{rel_fit_MOmegaI_eta_l2_rhs}}
  }
\end{table}

\begin{table}[H]
{\small  
  \centering
  \begin{tabular}{|c|c|c|c|c|c|c|}
    \hline
    \text{Theory} & $a$ & $b$ & $c$ & $d$ & $e$ & $\bar{\epsilon} [\%]$\\
    \hline
    \hline 
    GR &   $ 0.5 \pm 2.7$ 
    & $ -1 \pm 2.4 $ 
    &   $ 0.4 \pm 0.7 $ 
    &    $ 0.02 \pm 0.1$ 
    &  $ 0.086 \pm 0.005 $
    & $0.1$\\
    \hline 
      $m_{\phi}=0.343$neV &  $ -2 \pm 2$ 
    & $ 1 \pm 2$ 
    &   $ -0.33 \pm 0.65$ 
    &  $ 0.12 \pm 0.09$ 
    & $ 0.083 \pm 0.004 $
    & $0.1$\\
     \hline 
    $m_{\phi}=0.108$neV &  $  -3.7 \pm 3.7 $ 
    & $ 3.2 \pm 3.2 $ 
    &   $ -1.1 \pm 0.98 $ 
    &  $ 0.22 \pm 0.13 $ 
    & $  0.083 \pm 0.006  $
    & $0.2$\\
     \hline 
      $m_{\phi}=0.0343$neV &  $ 3 \pm 2 $ 
    & $ -2.8 \pm 1.7 $ 
    &   $ 0.86 \pm 0.53 $ 
    &  $ -0.07 \pm 0.07 $ 
    & $ 0.1 \pm 0.003 $
    & $0.1$\\
     \hline 
      $m_{\phi}=0.0108$neV &  $ 3 \pm 2$ 
    & $ -2.8 \pm 2 $ 
    &   $ 0.92 \pm 0.61 $ 
    &  $ -0.087 \pm 0.082 $ 
    & $0.102 \pm 0.004 $
    & $0.1$\\
     \hline 
  \end{tabular}
  {\caption{Fit parameters and their respective errors for Figure \ref{fig:uni_rel_omega_omegao} (left)} \label{rel_fit_omega_scale_hat_l2_lhs}}
  }
\end{table}

\begin{table}[H]
{\small  
  \centering
  \begin{tabular}{|c|c|c|c|c|c|c|}
    \hline
    \text{Theory} & $a$ & $b$ & $c$ & $d$ & $e$ & $\bar{\epsilon} [\%]$\\
    \hline
    \hline 
    GR &   $ 2.5 \pm 2$ 
    & $ -3.3 \pm 2.6 $ 
    &   $ 1.6 \pm 1.2 $ 
    &    $ -0.28 \pm 0.26$ 
    &  $ 0.11 \pm 0.02 $
    & $0.2$\\
    \hline 
      $m_{\phi}=0.343$neV &  $ 1.2 \pm 2.1$ 
    & $ -1.4 \pm 2.7$ 
    &   $ 0.6 \pm 1.3$ 
    &  $ -0.05 \pm 0.26$ 
    & $ 0.09 \pm 0.02 $
    & $0.2$\\
     \hline 
    $m_{\phi}=0.108$neV &  $  -0.1 \pm 2.2 $ 
    & $ 0.2 \pm 2.9 $ 
    &   $ -0.2 \pm 1.4 $ 
    &  $ 0.11 \pm 0.28 $ 
    & $  0.082 \pm 0.021  $
    & $0.2$\\
     \hline 
      $m_{\phi}=0.0343$neV &  $ 2.9 \pm 1.4 $ 
    & $ -3.8 \pm 1.8 $ 
    &   $ 1.8 \pm 0.8 $ 
    &  $ -0.33 \pm 0.17 $ 
    & $ 0.12 \pm 0.01 $
    & $0.1$\\
     \hline 
      $m_{\phi}=0.0108$neV &  $ 3 \pm 2$ 
    & $ -4 \pm 2 $ 
    &   $ 2 \pm 1 $ 
    &  $ -0.37 \pm 0.21 $ 
    & $0.12 \pm 0.02 $
    & $0.2$\\
     \hline 
  \end{tabular}
  {\caption{Fit parameters and their respective errors for Figure \ref{fig:uni_rel_omega_omegao} (right)} \label{rel_fit_omega_scale_hat_l2_rhs}}
  }
\end{table}

\begin{table}[H]
{\small  
  \centering
  \begin{tabular}{|c|c|c|c|c|c|c|}
    \hline
    \text{Theory} & $a\cdot10^3$ & $b\cdot10^3$ & $c\cdot10^3$ & $d\cdot10^3$ & $e\cdot10^3$ & $\bar{\epsilon} [\%]$\\
    \hline
    \hline 
    GR &   $ 5131 \pm 292$ 
    & $ -6858 \pm 383 $ 
    &   $ 3534 \pm 184 $ 
    &    $ -842 \pm 39$ 
    &  $ 82 \pm 3 $
    & $1.0$\\
    \hline 
      $m_{\phi}=0.343$neV &  $ 4907 \pm 283$ 
    & $ -6570 \pm 365$ 
    &   $ 3386 \pm 173$ 
    &  $ -805 \pm 36$ 
    & $ 78\pm 3 $
    & $0.9$\\
     \hline 
    $m_{\phi}=0.108$neV &  $  4327 \pm 336 $ 
    & $ -5707 \pm 431 $ 
    &   $ 2900 \pm 204 $ 
    &  $ -679 \pm 42 $ 
    & $  65 \pm 3  $
    & $0.9$\\
     \hline 
      $m_{\phi}=0.0343$neV &  $ 3575 \pm 163 $ 
    & $ -4645 \pm 208 $ 
    &   $ 2339 \pm 98 $ 
    &  $ -547 \pm 20 $ 
    & $  53 \pm 1 $
    & $0.6$\\
     \hline 
      $m_{\phi}=0.0108$neV &  $ 3664 \pm 149$ 
    & $ -4742 \pm 188 $ 
    &   $ 2373 \pm 88 $ 
    &  $ -550 \pm 18 $ 
    & $ 53 \pm 1 $
    & $0.5$\\
     \hline 
  \end{tabular}
  {\caption{Fit parameters and their respective errors for Figure \ref{fig:uni_rel_tau_omegao} (left)} \label{rel_fit_tau_scale_hat_l2_rhs}}
  }
\end{table}

\begin{table}[H]
{\small  
  \centering
  \begin{tabular}{|c|c|c|c|c|c|c|}
    \hline
    \text{Theory} & $a$ & $b$ & $c$ & $d$ & $e\cdot10^{-4}$ & $\bar{\epsilon} [\%]$\\
    \hline
    \hline 
    GR &   $ -0.1  \pm 0.02 $ 
    & $ 0.1 \pm 0.03 $ 
    &   $ -0.03 \pm 0.01 $ 
    &    $ 0.007 \pm 0.003$ 
    &  $ -4 \pm 2 $
    & $0.7$\\
    \hline 
      $m_{\phi}=0.343$neV &  $ -0.1 \pm 0.03 $ 
    & $ 0.1 \pm 0.04 $ 
    &   $ -0.04 \pm 0.02 $ 
    &  $ 0.01 \pm 0.004 $ 
    & $ -7 \pm 3 $
    & $ 0.8 $\\
     \hline 
    $m_{\phi}=0.108$neV &  $  -0.1 \pm 0.04 $ 
    & $ 0.1 \pm 0.05 $ 
    &   $ -0.04 \pm 0.02 $ 
    &  $ 0.01 \pm 0.005 $ 
    & $  -7 \pm 4  $
    & $0.8$\\
     \hline 
      $m_{\phi}=0.0343$neV &  $ -0.09 \pm 0.02 $ 
    & $ 0.06 \pm 0.03 $ 
    &   $ -0.02 \pm 0.01 $ 
    &  $ 0.005 \pm 0.003 $ 
    & $ -3 \pm 2 $
    & $0.5$\\
     \hline 
      $m_{\phi}=0.0108$neV &  $ -0.1 \pm 0.02$ 
    & $ 0.1 \pm 0.03 $ 
    &   $ -0.04 \pm 0.01 $ 
    &  $ 0.009 \pm 0.003 $ 
    & $ -6 \pm 2 $
    & $0.4$\\
     \hline 
  \end{tabular}
  {\caption{Fit parameters and their respective errors for Figure \ref{fig:uni_rel_tau_omegao} (right)} \label{rel_fit_tau_scale_hat_l2_lhs}}
  }
\end{table}

\begin{table}[H]
  \centering
  \begin{tabular}{|c|c|c|c|c|c|c|}
    \hline
    \text{Theory} & $a$ & $b$ & $c$ & $d$ & $e$ & $\bar{\epsilon} [\%]$\\
    \hline
    \hline 
    GR &   $ 8 \pm  3  $ 
    & $ -12 \pm 4  $ 
    &   $ 7 \pm 2  $ 
    &    $ -2 \pm 0.4  $ 
    &  $ 0.4 \pm 0.03 $
    & $0.2$\\
    \hline 
      $m_{\phi}=0.343$neV &  $ 5 \pm 3 $ 
    & $ -8 \pm 4 $ 
    &   $ 5 \pm 2 $ 
    &  $ -1 \pm 0.4 $ 
    & $ 0.3 \pm 0.03 $
    & $0.2$\\
     \hline 
    $m_{\phi}=0.108$neV &  $  3 \pm 4 $ 
    & $ -5 \pm 5 $ 
    &   $ 3 \pm 2 $ 
    &  $ -1 \pm 0.4 $ 
    & $  0.3 \pm 0.03  $
    & $0.2$\\
     \hline 
      $m_{\phi}=0.0343$neV &  $ 9 \pm 2 $ 
    & $ -13 \pm 3 $ 
    &   $ 8 \pm 1 $ 
    &  $ -2 \pm 0.3 $ 
    & $ 0.4 \pm 0.02 $
    & $0.1$\\
     \hline 
      $m_{\phi}=0.0108$neV &  $ 10 \pm 3 $ 
    & $ -15 \pm 3 $ 
    &   $ 8 \pm 2 $ 
    &  $ -2 \pm 0.3 $ 
    & $ 0.4 \pm 0.02 $
    & $0.2$\\
     \hline 
  \end{tabular}
  {\caption{Fit parameters and their respective errors for Figure \ref{fig:uni_rel_mass_tau__mass_omega_full} (left)} \label{rel_fit_R4_l2_rhs}}
\end{table}

\begin{table}[H]
  \centering
  \begin{tabular}{|c|c|c|c|c|c|c|}
    \hline
    \text{Theory} & $a$ & $b$ & $c$ & $d$ & $e\cdot10^{-5}$ & $\bar{\epsilon} [\%]$\\
    \hline
    \hline 
    GR &   $ -625 \pm  46  $ 
    & $ 22 \pm 3  $ 
    &   $ -0.15 \pm 0.08  $ 
    &    $ 0.0030 \pm 0.0008 $ 
    &  $ -1 \pm 0.3 $
    & $0.6$\\
    \hline 
      $m_{\phi}=0.343$neV &  $ -570 \pm 73 $ 
    & $ 18 \pm 5 $ 
    &   $ -0.05 \pm 0.1 $ 
    &  $ 0.002 \pm 0.001 $ 
    & $ -0.8 \pm 0.4 $
    & $0.6$\\
     \hline 
    $m_{\phi}=0.108$neV &  $  -614 \pm 85 $ 
    & $ 18 \pm 6 $ 
    &   $ -0.01 \pm 0.1 $ 
    &  $ 0.002 \pm 0.002 $ 
    & $  -0.8 \pm 0.5  $
    & $0.6$\\
     \hline 
      $m_{\phi}=0.0343$neV &  $ -692 \pm 70 $ 
    & $ 21 \pm 5 $ 
    &   $ -0.04 \pm 0.1 $ 
    &  $ 0.002 \pm 0.001 $ 
    & $ -0.9 \pm 0.4 $
    & $0.4$\\
     \hline 
      $m_{\phi}=0.0108$neV &  $ -819 \pm 67 $ 
    & $ 28 \pm 5 $ 
    &   $ -0.2 \pm 0.1 $ 
    &  $ 0.004 \pm 0.001  $ 
    & $ -1 \pm 0.4 $
    & $0.4$\\
     \hline 
  \end{tabular}
  {\caption{Fit parameters and their respective errors for Figure \ref{fig:uni_rel_mass_tau__mass_omega_full} (right)} \label{rel_fit_tau-OmegaR_l2_lhs}}
\end{table}

\begin{table}[H]
  \centering
  \begin{tabular}{|c|c|c|c|c|c|c|}
    \hline
    \text{Theory} & $a\cdot10^3$ & $b\cdot10^3$ & $c\cdot10^3$ & $d$ & $e$ & $\bar{\epsilon} [\%]$\\
    \hline
    \hline 
    GR &   $ 841 \pm 55 $ 
    & $ -786 \pm 48 $ 
    &   $ 288 \pm 15 $ 
    &    $ -49150 \pm 2041 $ 
    &  $ 3575 \pm 100 $
    & $1.1$\\
    \hline 
      $m_{\phi}=0.343$neV &  $ 822 \pm 91 $ 
    & $ -772 \pm 78 $ 
    &   $ 283 \pm 24 $ 
    &  $ -48342 \pm 3197 $ 
    & $ 3507 \pm 153 $
    & $1.3$\\
     \hline 
    $m_{\phi}=0.108$neV &  $  659 \pm 108 $ 
    & $ -618 \pm 92 $ 
    &   $ 228 \pm 29 $ 
    &  $ -39325 \pm 3835 $ 
    & $  2914 \pm 185  $
    & $1.3$\\
     \hline 
      $m_{\phi}=0.0343$neV &  $ 561 \pm 77 $ 
    & $ -528 \pm 66 $ 
    &   $ 197 \pm 21 $ 
    &  $ -34435 \pm 2794 $ 
    & $ 2606 \pm 135 $
    & $1.2$\\
     \hline 
      $m_{\phi}=0.0108$neV &  $ 622 \pm 76 $ 
    & $ -578 \pm 66 $ 
    &   $ 211 \pm 20 $ 
    &  $ -35943 \pm 2743 $ 
    & $ 2644 \pm 133 $
    & $1.2$\\
     \hline 
  \end{tabular}
  {\caption{Fit parameters and their respective errors for Figure \ref{fig:uni_rel_omega_tau_full} (left)} \label{rel_fit_tauOmegaR_l2_lhs}}
\end{table}

\begin{table}[H]
  \centering
  \begin{tabular}{|c|c|c|c|c|c|c|}
    \hline
    \text{Theory} & $a\cdot10^3$ & $b\cdot10^3$ & $c\cdot10^3$ & $d$ & $e$ & $\bar{\epsilon} [\%]$\\
    \hline
    \hline 
    GR &   $ 505 \pm  23  $ 
    & $ -664 \pm 30  $ 
    &   $ 336 \pm 14  $ 
    &    $ -78691 \pm 2997  $ 
    &  $ 7540 \pm 230 $
    & $0.8$\\
    \hline 
      $m_{\phi}=0.343$neV &  $ 471 \pm 23 $ 
    & $ -622 \pm 29 $ 
    &   $ 316 \pm 14 $ 
    &  $ -74140 \pm 2889 $ 
    & $ 7118 \pm 218 $
    & $0.7$\\
     \hline 
    $m_{\phi}=0.108$neV &  $  421 \pm 26 $ 
    & $ -549 \pm 34 $ 
    &   $ 276 \pm 16 $ 
    &  $ -64251 \pm 3318 $ 
    & $  6122 \pm 251  $
    & $0.7$\\
     \hline 
      $m_{\phi}=0.0343$neV &  $ 371 \pm 15 $ 
    & $ -478 \pm 19 $ 
    &   $ 239 \pm 9 $ 
    &  $ -55248 \pm 1810 $ 
    & $ 5292 \pm 136 $
    & $0.5$\\
     \hline 
      $m_{\phi}=0.0108$neV &  $ 384 \pm 14 $ 
    & $ -494 \pm 18 $ 
    &   $ 245 \pm 8 $ 
    &  $ -56241 \pm 1655 $ 
    & $ 5317 \pm 123 $
    & $0.4$\\
     \hline 
  \end{tabular}
  {\caption{Fit parameters and their respective errors for Figure \ref{fig:uni_rel_omega_tau_full} (right)} \label{rel_fit_tauOmegaR_eta_l2_rhs}}
\end{table}

\section{Comments and tables for universal relations for dipole F-mode}\label{App_dipole}

{This section starts with some historical comments on the construction of the dipole F-mode and then presents the tables of the fit parameters and their errors for the figures of the dipole F-mode universal relations in section \ref{Uni_rel_l1}.}

\begin{figure}[h!]
	\centering
	\includegraphics[width=.32\textwidth, angle =-90]{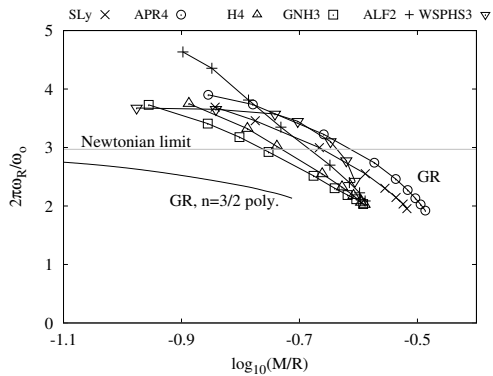}
	\caption{{F-mode: dimensionless frequency $\omega_R/ \omega_o$ versus logarithm of compactness $C=M/R$ for general relativity.
	The symbols indicate the respective EOS.
	The Newtonian limit is indicated by a thin horizontal line.
	}
	}
	\label{fig:uni_rel_c_l1_GRnPoly}
\end{figure}

{In order to check our analytical and numerical setup for the dipole F-mode, we have first reproduced some known results.}
Lindblom \cite{Lindblom:1989} has obtained the dipole F-mode for a $n=3/2$ polytrope EOS in general relativity and shown that it reaches the Newtonian limit for very small values of the compactness.
We have reproduced this result and show it in Figure \ref{fig:uni_rel_c_l1_GRnPoly}, where we show the real frequency $\omega_R$ scaled with the reference frequency, 
$\omega_o$ versus the logarithm of the compactness $C$.  
We recall that $\omega_o = \sqrt{\frac{3G \tilde M}{4R^3}}=\frac{c}{M}\sqrt{\frac{3}{4}C^3}$, which involves both the mass $\tilde M$ ($M=G \tilde M/c^2$) and the radius $R$ of the neutron star in the combination of its mean density.
We have marked the Newtonian limit of the polytrope ($\omega_R = 2.9696 \omega_o$) with a horizontal line.

In Figure \ref{fig:uni_rel_c_l1_GRnPoly} we show also our results for the here employed realistic EOSs in general relativity. 
For these EOSs we obtain somewhat larger values than for the  polytrope for the same compactness, but we have also considered neutron stars with larger values of the compactness. 
In fact the polytrope used in \cite{Lindblom:1989} does not produce neutron stars with sufficiently high masses and is therefore observationally excluded.
{Since the curves for the various EOSs remain rather distinct in this figure, this indicates already that use of the dimensionless frequency $\omega_R/ \omega_o$ versus the logarithm of the compactness $C=M/R$ will not yield a convincing universal relation for general relativity.}

{In the following we exhibit the tables of the fit parameters and their errors for the figures of the dipole F-mode universal relations in section \ref{Uni_rel_l1}.}

\begin{table}[H]
  \centering
  \begin{tabular}{|c|c|c|}
    \hline
    \text{Theory} & $c$ & $\bar{\epsilon} [\%]$\\
    \hline
    \hline 
    $m_{\phi}=0.1084$neV &  $3.96\pm 0.02$ & $3.3$\\
     \hline 
    $m_{\phi}=0.0766$neV  & $2.88\pm 0.004$ & $0.8$\\
     \hline 
    $m_{\phi}=0.0343$neV  & $1.295\pm 0.001$ & $0.4$\\
     \hline 
  \end{tabular}
  {\caption{Fit parameters and their respective errors for Figure \ref{fig:analysis_l1} } \label{rel_fit_analysis_l1_lhs}}
\end{table}

\begin{table}[H]
  \centering
  \begin{tabular}{|c|c|c|c|}
    \hline
    \text{Theory} & $b$ & $c$ & $\bar{\epsilon} [\%]$\\
    \hline
    \hline 
    GR &   $0.13 \pm 0.01$ & $0.010 \pm 0.002$ & $9.4$\\
    \hline 
    $m_{\phi}=0.343$neV &  $0.12\pm 0.01$ & $0.011\pm 0.002$ & $9.1$\\
     \hline 
    $m_{\phi}=0.108$neV  & $0.127\pm 0.006$ & $0.007\pm 0.001$ & $5.2$\\
     \hline 
    $m_{\phi}=0.0767$neV  & $0.094\pm 0.005$ & $0.005\pm 0.001$ & $5.7$\\ 
     \hline 
    $m_{\phi}=0.0343$neV  & $0.044\pm 0.002$ & $0.002\pm 0.0006$ & $6.1$\\
     \hline 
  \end{tabular}
  {\caption{Fit parameters and their respective errors for Figure \ref{fig:uni_rel_MOmegaR_error_eta_linear_affine} (left)} \label{rel_fit_MOmegaR_lhs}}
\end{table}

\begin{table}[H]
  \centering
  \begin{tabular}{|c|c|c|}
    \hline
    \text{Theory} & $b$ & $\bar{\epsilon} [\%]$\\
    \hline
    \hline 
    GR &   $0.110 \pm 0.002$ & $10.4$\\
    \hline 
    $m_{\phi}=0.343$neV &  $0.111\pm 0.002$ & $10.3$\\
     \hline 
    $m_{\phi}=0.108$neV  & $0.104\pm 0.001$ & $5.0$\\
     \hline 
    $m_{\phi}=0.0343$neV  & $0.0353\pm 0.0003$ & $5.1$\\
     \hline 
  \end{tabular}
  {\caption{Fit parameters and their respective errors for Figure \ref{fig:uni_rel_MOmegaR_error_eta_linear_affine} (right)} \label{rel_fit_MOmegaR_rhs}}
\end{table}

\begin{table}[H]
  \centering
  \begin{tabular}{|c|c|c|c|}
    \hline
    \text{Theory} & $b$ & $c$ & $\bar{\epsilon} [\%]$\\
    \hline
    \hline 
    GR &   $-1.7 \pm 0.1 $ & $0.82 \pm 0.03$ & $10.1$\\
    \hline 
    $m_{\phi}=0.343$neV &  $-1.9\pm 0.1 $ & $0.87 \pm 0.03$ & $9.9$\\
     \hline 
    $m_{\phi}=0.108$neV  & $-1.53\pm 0.07 $ & $0.76\pm 0.02$ & $5.3$\\
     \hline 
    $m_{\phi}=0.0767$neV  & $-1.08 \pm 0.06$ & $0.55\pm 0.01$ & $4.5$\\ 
     \hline 
    $m_{\phi}=0.0343$neV  & $-0.44\pm 0.03$ & $0.241\pm 0.006$ & $4.5$\\
     \hline 
  \end{tabular}
  {\caption{Fit parameters and their respective errors for Figure \ref{fig:uni_rel_c_2_l1_error_no_log_linear} (left)} \label{rel_fit_Omegao_MR}}
\end{table}

\begin{table}[H]
  \centering
  \begin{tabular}{|c|c|c|c|}
    \hline
    \text{Theory} & $b$ & $c$ & $\bar{\epsilon} [\%]$\\
    \hline
    \hline 
    GR &   $-1.46 \pm 0.09$ & $0.95 \pm 0.03$ & $8.1$\\
    \hline 
    $m_{\phi}=0.343$neV &  $-1.58\pm 0.09$ & $1 \pm 0.03$ & $7.9$\\
     \hline 
    $m_{\phi}=0.108$neV  & $-1.26 \pm 0.05$ & $0.85\pm 0.02$ & $4.2$\\
     \hline 
    $m_{\phi}=0.0343$neV  & $-0.37\pm 0.02$ & $0.266\pm 0.008$ & $5.6$\\
     \hline 
  \end{tabular}
  {\caption{Fit parameters and their respective errors for Figure \ref{fig:uni_rel_c_2_l1_error_no_log_linear} (right)} \label{rel_fit_Omegao_rhs}}
\end{table}

\newpage

\bibliographystyle{unsrt}

\end{document}